\documentclass[12pt]{iopart}  

\usepackage{graphicx} 
\usepackage{bm} 

\expandafter\let\csname equation*\endcsname\relax
\expandafter\let\csname endequation*\endcsname\relax
\usepackage{amsmath}

\usepackage{epsfig}
\usepackage{color}
\usepackage{dcolumn} 

\newcommand{\ket}[1]{|#1\rangle}
\newcommand{\bra}[1]{\langle#1|}

\newcommand{\beq}{\begin{eqnarray*}}
\newcommand{\eeq}{\end{eqnarray*}}

\newcommand{\red}[1]{{#1}}

\begin{document}

\title{Justifying quasiparticle self-consistent schemes {\it via} gradient optimization in \red{Baym-Kadanoff} theory}

\author{Sohrab Ismail-Beigi} 
\address{Department of Applied Physics,
Department of Physics,
Department of Mechanical Engineering and Materials Science, and
Center for Research on Interface Structures and Phenomena,\\
Yale University, New Haven, CT 06520, U.S.A.}

\begin{abstract}
The question of which non-interacting Green's function ``best'' describes an interacting many-body electronic system is both of fundamental interest as well as of practical importance in describing electronic properties of materials in a realistic manner.  Here, we study this question within the framework of \red{Baym-Kadanoff} theory, an approach where one locates the stationary point of a total energy functional of the one-particle Green's function in order to find the total ground-state energy as well as all one-particle properties such as the density matrix, chemical potential, or the quasiparticle energy spectrum and quasiparticle wave functions.  \red{For the case of the Klein functional}, our basic finding is that minimizing the length of the gradient of the total energy functional over non-interacting Green's functions yields a set of self-consistent equations for quasiparticles that is identical to those of the Quasiparticle Self-Consistent $GW$ (QS$GW$) \cite{QPSCGW} approach, thereby providing an {\it a priori} justification for such an approach to electronic structure calculations.  In fact, this result is general, applies to any self-energy operator, and is not restricted to any particular approximation, {\it e.g.} the $GW$ approximation for the self-energy.  The approach also shows that, when working in the basis of quasiparticle states, solving the diagonal part of the self-consistent Dyson equation is of primary importance while the off-diagonals are of secondary importance, a common observation in the electronic structure literature of self-energy calculations.  Finally, numerical tests and analytical arguments show that when the Dyson equation produces multiple quasiparticle solutions corresponding to a single non-interacting state,  minimizing the length of the gradient   translates into choosing the solution with largest quasiparticle weight.
\end{abstract}

\submitto{\JPCM}
\maketitle

\section{Introduction}
\label{sec:intro}

Single-particle approaches for computing the electronic structure of materials have proven very useful   for understanding and predicting the properties of materials, particularly {\it ab initio} methods such as Density Functional Theory (DFT) \cite{HK,KS}.  The local density (LDA) or generalized gradient (GGA) approximations \cite{KS,LDA,GGA} for DFT provide practical computational approaches that are the {\it de facto} workhorses for obtaining total energies, atomic geometries, vibrational modes, thermodynamic data, chemical properties, kinetic barriers, {\it etc.} of a great variety of materials.  Aside from practical usefulness, the single-particle nature of these approaches permits one to straightforwardly analyze the link between the atomic-scale structure of the material and the resulting electronic structure, {\it e.g.}, {\it via} tight-binding or nearly free-electron models.  The relative straightforwardness of a single-particle framework permits one to then propose materials design principles whereby one can tune or engineer desirable materials properties.  Nevertheless, there are some shortcomings to such an approach.  One can categorize the main drawbacks of single-particle schemes such as DFT for electronic structure predictions into two broad categories.  

The first is fundamental to the single-particle approach itself when it is applied to strongly correlated electronic systems.  When the basic behavior of electrons is determined by strong and localized electronic repulsions, it is  difficult to properly describe such a situation using single-particle approaches where each particle moves separately in an effective potential \cite{DMFT1,DMFT2}.  A number of methods have been proposed to date to deal with such situations, and at present Dynamical Mean Field Theory \cite{DMFT1,DMFT2} represents a workable scheme with the requisite compromise between reasonable computational complexity (obtained by approximating the many-body correlated problem in certain ways) and realistic description of actual materials.   Even in such cases, however, building a many-body description of the correlated system in a method such as DMFT requires inclusion of important single-particle terms that reflect the structure, local chemistry and bonding; the strong interactions are added on top of this, as exemplified by the canonical Hubbard model and its various extensions.  Thus one needs an ``optimal'' single-particle description to begin the process.

A second drawback is due to the ground-state nature of DFT approaches and the use of a local effective potential: even without strong correlations, a theory designed to describe the ground state with a local potential will have a difficult time predicting excited state properties such as band energies and band gaps \cite{Perdewgapproblem,ShamSchluter,LDAbadgaps}.  In a number of cases, one can correct the main faults with self-interaction corrected approaches \cite{LDA} or explicit inclusion of a degree of Fock exchange in hybrid approaches \cite{hybrid1,hybrid2,hybrid3}.  The popular LDA+U approach~\cite{LDApU} falls into this category where Fock-type corrections are included for a subspace of states spanned by pre-chosen localized atomic-like orbitals.  The main idea in all these methods is to add more complexity to the effective potential in order to better incorporate the important physics of Fock exchange and to remove the closely related problem of electron self-interaction that plagues the canonical DFT approximations.  A more {\it ab initio} approach that does not require pre-determined localized basis sets or pre-chosen physical effects is to use the many-body perturbation theory of Green's functions.  The most successful to date is the $GW$ approximation to the self-energy~\cite{Hedin,strinati_dynamical_1980,strinati_dynamical_1982,HL,TDDFTvsGWBSE}.  The $GW$ approximation delivers high quality band structures of many band insulators and simple metals and automatically includes many physical effects such as exact Fock exchange, localized Coulomb repulsion, dynamic screening, and dispersion forces.  In addition, LDA+U is a static and localized approximation to $GW$~\cite{LDApU}, and the effective potentials used in hybrid methods generally include a subset of the physics in $GW$ (mainly Fock exchange that is screened in some manner) \cite{hybrid1,hybrid2,hybrid3}.  Most $GW$ calculations are performed perturbatively: they compute corrections to an input DFT-like electronic structure.  The final result in turn depends on the input description: in cases where the LDA provides a decent starting point, $GW$ corrections provide a good description of the electronic structure \cite{GWNiO,TMOGWmodel,GWZnO,GWMgCaTiVO,GWaeSiMnONiO,GWNiO2}.  But in other situations, the inadequacy of the input DFT description can create quantitative errors \cite{GWaeSiMnONiO,QPSCGW,scgwCu2O,scCOHSEX}.  

Ideally, one would like to overcome the starting-point dependence by doing a self-consistent calculation within the $GW$ approach itself.  One would aim to have an approach that does not assume any particular basis set or rely on some set of parameters.  Among many possibilities, two methods have been used by a number of researchers.  One is the Quasiparticle Self-Consistent $GW$ (QS$GW$) \cite{QPSCGW}, and the other is the self-consistent COHSEX (scCOHSEX) \cite{scCOHSEX}.  Both move one away from having to use DFT as the starting point for a $GW$ calculation by finding a non-interacting band structure approximately but self-consistently.  What this means is that one has a parameter-free method to automatically include static and dynamic screening, Fock exchange, certain aspects of localized Coulombic physics in a single calculation.  

While these methods represent exciting developments, they are based on physical insight and/or approximation of the $GW$ self-energy operator to yield workable schemes.  A key question is if there is some theoretical sense in which one can derive an optimal non-interacting band structure for any electronic system, and what such a description would look like.  Namely, do these schemes, or various modifications of them, have an {\it a priori} theoretical justification?  

In this work, we answer this question positively by showing that within the appropriate total-energy scheme appropriate for Green's function methods, namely, the \red{Baym-Kadanoff approach 
\cite{baym_conservation_1961,baym_self-consistent_1962} together with the Klein functional \cite{Klein}}, quasiparticle self-consistent approaches are the most theoretically justified in the sense that they are ``closest'' to the stationary point of the energy functional.  The quantitative meaning of closeness is based on gradient minimization, namely the length of the gradient of the energy functional is minimized within the search space considered.  Our results thus justify the use of the QS$GW$ approach.  We emphasize, however, that our main result is not only applicable to  the $GW$ approximation alone but to any self-energy operator. The QS scheme is the optimal one for generating a non-interacting band structure (in the sense of gradient optimization).

The remainder of this paper is organized as follows. Section~\ref{sec:defs} describes our notation and definitions.  Section~\ref{sec:lw} reviews the \red{Baym-Kadanoff} approach specifically within the framework of finding optimal non-interacting band structure.  Section~\ref{sec:eta} describes the small parameter used to organize the gradient optimization process.  Section~\ref{sec:gradient} describes how gradient optimization leads to quasiparticle self-consistent equations.  Section~\ref{sec:deltaF} describes an alternative approach for optimization that one might consider based on optimizing the value of the energy functional itself, but it is shown that this approach, while tempting and easy to state, does not seem to lead to further insights or useful results.  Section~\ref{sec:application} shows, by numerical solving a simple many-body system as well as by providing more general arguments, that the gradient optimization approach requires choosing the quasiparticle solution with largest weight ($Z$) when deciding among multiple solutions to the Dyson equation that all correspond to a single non-interacting state.  Section~\ref{sec:conclusions} has a brief summary of  results and their implications.

\section{Definitions \& Notation}
\label{sec:defs}

We assume our electronic system is governed by a time-independent and time-reversal invariant many-body Hamiltonian which means that many key physical quantities such as wave functions are real-valued and all time-dependent quantities depend only on time differences.  We use atomic units so $\hbar=1$, the unit of elementary charge $e=1$, and the electron mass $m_e=1$.  Hence, energies and frequencies are interchangeable.  Wherever sensible, we use matrix notation for compactness.  For example, the one-particle electron Green's function for a time-independent system in the frequency $\omega$ domain, $G(x,x',\omega)$, is a function of three arguments.  The $x$ and $x'$ arguments include both spatial coordinates and spin: $x=(\vec r, \sigma)$ where $\vec r$ is a three-vector and $\sigma=\pm 1$ labels the two spin projections.  In matrix notation, we write the matrix $G(\omega)$ whose matrix elements are $\bra{x}G(\omega)\ket{x'}=G(x,x',\omega)$.  

We begin with the non-interacting system.  It is specified by  the chemical potential $\mu$ and the static (frequency-independent) Hermitian single-particle Hamiltonian $H_0$ with orthonormal eigenstates $\ket{n}$ and real eigenvalues $\epsilon_n$
\[
H_0\ket{n} = \epsilon_n \ket{n}\,.
\]
The unoccupied (conduction) states have $\epsilon_n>\mu$ and the occupied (valence) states have $\epsilon_n<\mu$.  For direct comparison to the interacting Green's function and its self-energy, we separate from $H_0$ the part that deals with exchange and correlation $U_{xc}$.  We define
\[
H_0 = T + U_{ion} + \phi_H + U_{xc}\,.
\]
Here, $T=-\nabla^2/2$ is the electron kinetic energy operator,  $U_{ion}$ is the electron-ion interaction potential, the Hartree potential $\phi_H$ is determined by the electron density $n(x)$ {\it via}
\[
n(x) = \sum_n \theta(\mu-\epsilon_n)|\psi_n(x)|^2
\]
{\it via}
\[
\phi_H(x) =  \sum_{\sigma'} \int dr'\ \frac{n(x')}{||\vec r-\vec r\,'||}\,.
\]
The static and Hermitian $U_{xc}$ aims to approximate the exchange and correlation effects.  In DFT, $U_{xc}$ is a local potential.  Here, we consider a more general   non-local $U_{xc}$, {\it i.e.}, $U_{xc}(x,x')\ne0$ when $x\ne x'$. 

The frequency domain non-interacting Green's function $G_0(\omega)$ is given by
\[
G_0(\omega)^{-1} = \omega I - H_0 = \omega I - T - U_{ion} - \phi_H - U_{xc}\,.
\]
We note that for a fixed nuclear configuration and thus fixed $U_{ion}$, $U_{xc}$ determines $G_0$ and {\it vice versa}.  This is a useful parameterization of $G_0$ that we will see below.  Finally, the non-interacting one-particle density matrix $\rho_0$ is given by
\[
\rho_0(x,x') = \sum_n \theta(\mu-\epsilon_n) \psi_n(x)\psi_n(x')^*
\]
whose diagonal is the density $n(x)=\rho_0(x,x')$.

In the frequency domain, the interacting Green's function is given  compactly by the Dyson equation
\begin{equation}
G(\omega)^{-1} = \omega I - T - U_{ion} - \phi_H - \Sigma_{xc}(\omega)
\label{eq:traddyson}
\end{equation}
 and the self-energy $\Sigma_{xc}(\omega)$ is frequency-dependent (dynamic) and non-Hermitian and encodes the complex exchange and correlation effects of the many-body system.  For use in this article, the Dyson equation can be rewritten as
\begin{equation}
G^{-1}(\omega) = G_0^{-1}(\omega) - \Big[\Sigma_{xc}(\omega) - U_{xc}\Big]\,.
\label{eq:dysoneq}
\end{equation}
Therefore, to find the true interacting Green's function, we must replace the static, Hermitian $U_{xc}$ by the dynamic, non-Hermitian $\Sigma_{xc}(\omega)$.    

The interacting density matrix $\rho$ is most compactly specified by taking the zero-time value of time-domain Green's function $\mathcal{G}(x,x',t)$
\[
\rho(x,x') = -i\mathcal{G}(x,x',t=0^-)\,.
\]
Here, $\mathcal{G}(t)$ is the Fourier transform of $G(\omega)$ and  $0^-$ is a negative infinitesimal.  This density matrix appears in the expression for the Fock exchange operator.

Finally, for immediate use below, we define two trace operators.  For any matrix $A$, we let $tr\{A\}$ denote the standard definition
\[
tr\{A\} \equiv \int dx \, A(x,x) = \sum_n \bra{n}A\ket{n}\,.
\]
Given a matrix that is a function of frequency, $B(\omega)$, we define the shorthand $Tr\{B\}$ to stand for the integral
\[
Tr\{B\} \equiv \int_{-\infty}^\infty \frac{d\omega}{2\pi i} e^{i\omega0^+}tr\{B(\omega)\}
\]
where one can convert the integral to a closed contour integral by going over the upper complex $\omega$ plane by using the convergence factor $e^{i\omega0^+}$ where $0^+$ is a positive infinitesimal.

\section{Baym-Kadanoff approach}
\label{sec:lw}

The basic point of the  approach of \red{Baym-Kadanoff} \cite{baym_conservation_1961,baym_self-consistent_1962} is that both the ground-state total energy and the interacting Green's function $G(\omega)$ can be obtained by finding the stationary point of an energy functional of $G$.  For simplicity, we  concentrate on the Klein functional \cite{Klein}, a functional of both $G$ and the  non-interacting $G_0$ given by
\begin{multline}
F[G,G_0] = E_H[n] + \Phi_{xc}[G] + \\
Tr\Big\{ H_0G_0 + I - G_0^{-1}G + \ln[G_0^{-1}G] -(\phi_H^0+U_{xc})G\Big\} \,.
\label{eq:kleinfunc}
\end{multline}
where $\phi^0_H$ is the Hartree potential for the electron density corresponding to $G_0$.
In the above expression, the frequency dependence of $G(\omega)$ and $G_0(\omega)$ has been suppressed for clarity.
$E_H[n]$ is the Hartree energy for electron density $n(x)$:
\[
E_H[n] =  \frac{1}{2} \int dx \ n(x) \phi_H(x)\,.
\]
The functional $\Phi_{xc}[G]$ is the exchange-correlation energy functional for the Baym-Kadanoff approach and, as in DFT, is a complicated and unknown functional of $G$.  Formally, it has a well-defined diagrammatic expansion~\cite{LW}.  
As in DFT, choosing an approximate form for $\Phi_{xc}$ corresponds to including a certain approximate level of treatment of exchange-correlation effects.  \red{The Klein functional is not the only possible variational functional in Baym-Kadanoff theory: the Luttinger-Ward functional \cite{LW} is more widely known, is known to have better variational properties~\cite{DahlenvonBarthPRB04} but has a much more complex form.  Hence, we will be focusing on the simpler Klein functional.}

At the stationary point of $F$ \red{(for both the Klein and Luttinger-Ward forms)}, the value of $F$ is the ground-state total energy, and the stationary $G$ is the true one-particle Green's function \cite{LW,Hedin}.  Unlike DFT, one can obtain, in principle, not just the ground state total energy and electron density but also excited state properties such as quasiparticle wave functions and band energies.  Much like the Kohn-Sham DFT approach, there is a functional derivative relation between the exchange-correlation energy functional and the self-energy
\begin{equation}
\Sigma_{xc}(\omega) = 2\pi i \frac{\delta \Phi_{xc}[G]}{\delta G(\omega)}\,.
\label{eq:sigdefphixcderiv}
\end{equation}

To make practical progress in the \red{Baym-Kadanoff} framework, two separate types approximations are necessary.  The first is the same as that encountered in DFT: one must choose some approximate $\Phi_{xc}$.  The second challenge is that, unlike DFT where $N$-presentability conditions are known \cite{Gilbert,Harriman}, similar conditions for the Green's function $G(x,x',\omega)$ are unknown.  Namely, one does not know which subset of functions $G(x,x',\omega)$ correspond to physically realizable Greens functions for the standard interacting electronic many-body Hamiltonian.  Therefore, one can try to directly tabulate and work with an arbitrary function $G(x,x',\omega)$ to locate the stationary point of $F$, which will hopefully correspond to the physical $G$, but such an approach is very demanding and computationally expensive.  Alternatively, one can make some simplifying assumptions on the types of Green's functions considered.

Here, we restrict ourselves to using non-interacting Green's functions for $G$.  Since it is known that $F$ does not in fact depend on $G_0$ (for fixed $G$)~\cite{myPRB}, once we restrict $G$ to be non-interacting, we might as well set $G_0$ equal to $G$ for convenience without any change to $F$.    This also simplifies the functional significantly to
\begin{equation}
F[G_0,G_0] = tr \Big\{ [T+U_{ion}] \rho_0 \Big\} + E_H[n]+ \Phi_{xc}[G_0]\,.
\label{eq:FG0G0}
\end{equation}
This energy functional contains  familiar  terms: the noninteracting kinetic, electron-ion, and Hartree energy plus the exchange-correlation contribution.  The first three terms are identical to their DFT counterparts and depend only on $\rho_0$.  Only $\Phi_{xc}$ depends on the added dynamical information in the Green's function $G_0$.
\begin{figure}[t!]
\includegraphics[width=3in]{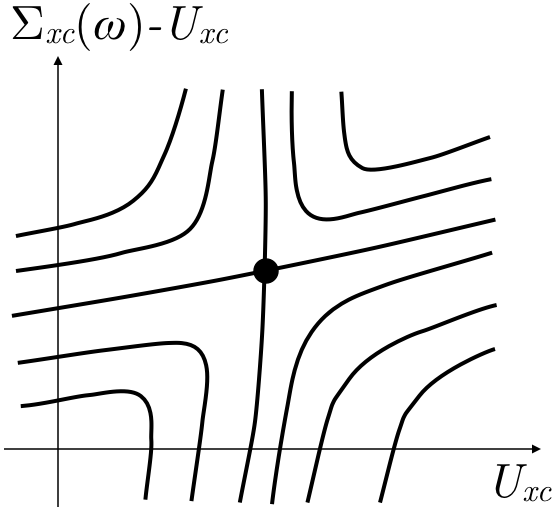}
\caption{A schematic of the simplest scenario for the \red{Baym-Kadanoff} energy functional  $F$.  The thick solid lines are level curves of $F$.  The self-energy $\Sigma_{xc}(\omega)$ that parameterizes the trial Green's function $G(\omega)$ is divided into two distinct contributions:  a static and Hermitian part $U_{xc}$ that parameterizes the non-interacting Green's functions $G_0$, and a remaining dynamical, non-Hermitian part $\Sigma_{xc}(\omega)-U_{xc}$.  These two independent contributions are high-dimensional matrices but as schematically shown as independent axes.  The black circle represents the stationary point of the energy functional which corresponds to the true  self-energy that self-consistently solves the Dyson equation (\ref{eq:dysoneq}).  The horizontal axis represents the space of all non-interacting Green's functions.  We see that the level curves cross this axis with no interruption reflecting the known fact that $F$ has no stationary point when sampled along the horizontal axis~\cite{myPRB}.  As the Figure illustrates, this also implies that the stationary point of $F$ must be a saddle point.
}
\label{fig:contours}
\end{figure}

Due to the stationary nature of $F$ about the optimal $G$, $F[G_0,G_0]$ provides a variational estimate of the ground-state energy with the error being smallest for the ``best'' $G_0$.  The main problem, which we have tried to address previously~\cite{myPRB} and which we address in this work, is how to choose a best $G_0$ and what this means.  A tempting idea is to try to minimize or optimize $F[G_0,G_0]$ over various trial $G_0$ or equivalently over various trial $U_{xc}$.  However, this program is highly problematic because $F[G_0,G_0]$ does not have any stationary points~\cite{myPRB}.  Figure \ref{fig:contours} illustrates the situation graphically and schematically.  The stationary point of $F$ representing the true Green's function that solves the Dyson equation (\ref{eq:dysoneq}) must correspond to a saddle point and to a dynamic and non-Hermitian self-energy, whereas if we constrain ourselves to static and Hermitian self-energies $U_{xc}$, then $F$ has non-zero derivatives in the whole subspace.

We discuss three avenues for avoiding this pathological situation.  The first is to constrain the types of $G_0$ or $U_{xc}$ being considered based on physical knowledge or intuition.  For example, it is known that forcing $U_{xc}$ to be a local potential is sufficient  to create a minimum for the Klein functional~\cite{DahlenvonBarthPRB04, DahlenLeeuwenvonBarth06, HellgrenvonBarth07}.  However, a local potential can not directly and consistently describe a number of simple non-local effects, {\it e.g.}, Fock exchange which  automatically removes problematic electron self-interaction effects.  Physically motivated non-local forms for $U_{xc}$ are exemplified by the QS$GW$ or scCOHSEX approaches which are based on incorporating $GW$-level self-energy effects into $U_{xc}$.

A second idea is to recast the optimization process by working with a different variable.  For example, an interesting recent work \cite{blochl_density-matrix_2013} shows that, in principle, one can optimize the \red{Baym-Kadanoff} energy functional by  using the density matrix $\rho$ as the fundamental variable (instead of $G$) and have a minimum principle for the resulting energy functional.  The main difficulty is that one can not avoid the fact that the stationary point of the \red{Baym-Kadanoff} functional is a saddle point, so that the  optimization process involving $\rho$ still requires an internal search at fixed $\rho$ for a saddle point \cite{blochl_density-matrix_2013}.  To avoid this complexity, an ``NDE2'' approximation has been proposed \cite{blochl_density-matrix_2013} which removes this saddle point search: it basically consists of evaluating $\Phi_{xc}$ at $G_0$ instead of at $G$ (very similar in spirit to Eq.~(\ref{eq:FG0G0})).  It is our belief that the NDE2  will  suffer from this same problems discussed above: $\Phi_{xc}[G_0]$ has no lower bound and the minimization will drive the system to an unphysical minimum with negative infinite energy \cite{myPRB}.  However, future studies are needed to carefully evaluate this matter.  Overall, the density matrix approach to \red{Baym-Kadanoff} theory is a promising new idea in the field.

A third approach is to come up with a different mathematical definition of the ``best'' $G_0$ which does not rely on na\"ive  optimization of $F$.  We follow this more mathematical approach which will  yield a self-consistent quasiparticle scheme.  Instead of trying to satisfy the impossible condition $\delta F/\delta G_0=0$, we will minimize the length of the gradient of $F$.  Specifically, we minimize the square length of the gradient $|\delta F/\delta \Sigma_t|^2$ where $\Sigma_t$ is a trial self-energy.  Our results will be general as we will not assume any specific   approximation for the self-energy so that the main results will hold for any chosen form of   $\Phi_{xc}$.  Our approach has similarities to the Optimized Effect Potential (OEP) method~\cite{sharp_variational_1953, talman_optimized_1976, kummel_orbital-dependent_2008} in DFT.  In OEP, one finds the optimal local potential corresponding to the minimum of some general and possibly orbital dependent exchange-correlation energy functional $E_{xc}$: one has a formalism to find the self-consistent local Kohn-Sham potential corresponding to the total energy functional.  Hence, both approaches are looking for an optimized and simplified single-particle description (a local potential in OEP and a non-local $H_0$ in \red{Baym-Kadanoff} theory).  However,    whereas in OEP one can find a local potential that minimizes the total energy functional so the functional derivative of the total energy versus the potential is zero, in the \red{Baym-Kadanoff} setting one must settle for a non-zero derivative so that the structure and interpretation of the resulting equations for the non-local single-particle Hamiltonian are more complex.

\section{Key small parameter {$\gamma$}}
\label{sec:eta}

As explained above, this paper is  focused primarily on the use of non-interacting Green's functions $G_0$.  We  will be working primarily in  frequency space, which is the natural representation when discussing quasiparticle energies and how to deal with the frequency dependence of $\Sigma_{xc}(\omega)$.  In what follows, we will be employing time-ordered non-interacting Green's functions $G_0(\omega)$ of the following form:
\begin{equation}
G_0(\omega) = \sum_n \frac{\ket{n}\bra{n}}{\omega-\epsilon_n- i\gamma s_n }
\label{eq:G0omega}
\end{equation}
where $s_n = \mbox{sgn}(\mu-\epsilon_n)$\,.  We choose the real-valued quantity $\gamma>0$ to be small and fixed.   Note that we are {\it choosing} to use this class of $G_0$.

In standard textbook treatments~\cite{negele_quantum_1998,fetter_quantum_2003}, one finds the form of Eq.~(\ref{eq:G0omega}) where the symbol $\eta$ takes the place of $\gamma$, and $\eta\rightarrow0^+$ is understood or imposed at some point.  The quantity $\eta$ in these standard treatments is a mathematical quantity: an infinitesimal positive that is needed to ensure that Fourier transformations are absolutely convergent when Fourier transforming from time to frequency domain.  Often, it is directly included or absorbed into the definition of the Heaviside $\theta$ function for the time-ordering operation~\cite{fetter_quantum_2003}.  

In our case, however, $\gamma$ is positive and small but always finite.  We do {\it not} send it to zero.  It is a fixed number imposed by us manually with a simple physical meaning: it is a damping rate or inverse quasiparticle lifetime for the $\epsilon_n$ energy states.  We are giving our non-interacting states in our input Green's function $G_0$ of Eq.~(\ref{eq:G0omega}) a finite but very small decay rate $\gamma$ which is equivalent to a very large but finite lifetime $\gamma^{-1}$.  

Mathematically, we will see that $\gamma$ acts as a small parameter  that (i) regularizes the mathematical expressions we compute by ensuring that they are finite ({\it i.e.}, avoiding division by zero in energy denominators), and (ii) permits rank-ordering of dominant versus subdominant contributions.  Hence, from a mathematical viewpoint, it is necessary to keep $\gamma$ finite.

However, from a {\it physical} viewpoint, the finite value of $\gamma$ is hardly a restriction.  We can take $\gamma$ to be very small so that the imposed lifetime $\gamma^{-1}$ can be quite long (a day, a month, a year) and certainly longer than any contemplated experiment which would measure the electronic response.  Physically, we need to make $\gamma$ small enough to resolve 
and not  spoil intrinsic energetic shifts, broadenings and lifetimes stemming from electron interactions and scattering.  For example, if the system has a quasiparticle energy gap $\Delta$, then $\gamma\ll \Delta$ is required to resolve this gap precisely.  Or, if a low-energy quasiparticle state has a lifetime $\tau$ due to electron-electron scattering, then $\gamma\ll 1/\tau$ is needed in order to correctly obtain this lifetime {\it via} a calculation of the Green's function.  Finally, turning on electron-electron interactions modifies the spectrum of the Green's function (the spectral function) away from its non-interacting analogue.  As per Eq.~(\ref{eq:dysoneq}), the energy scale of such changes is determined by the magnitude of the matrix elements of $\Sigma_{xc}-U_{xc}$, so we require $\gamma$ to be smaller than these matrix elements to ensure well-converged spectra.  This last requirement is directly reflected in the key equations of our analysis below: ratios of matrix elements to $\gamma$
\[
\frac{\bra{n}\Sigma_{xc}-U_{xc}\ket{m}}{\gamma}
\]
appear below and represent the large quantities that must be minimized to obtain the optimal $U_{xc}$.

\section{Shortest Gradient of $F$}
\label{sec:gradient}

In this section, we implement the standard idea of minimizing the length of the gradient vector: the smaller the gradient of $F$, the closer we should be to the stationary point.  Specifically, we seek the optimum non-interacting $G_0$ that delivers the shortest gradient of $F$.

We begin with the following expression for the variation of $F$ versus $G$ (for fixed $G_0$ and arbitrary $G$) that is  derived by differentiating Eq.~(\ref{eq:kleinfunc}):
\begin{eqnarray}
\delta F & = & Tr \Big\{\left[
G^{-1}-G_0^{-1} + \Sigma_{xc}-U_{xc} \right] \delta G
\Big\}\,.
\label{eq:dFdG}
\end{eqnarray}
As a reminder, $\Sigma_{xc}=2\pi i\delta \Phi_{xc}/\delta G$.
As a matrix derivative, this is equivalent to
\begin{equation}
2\pi i\frac{\delta F}{\delta G(\omega)} = G(\omega)^{-1}-G_0(\omega)^{-1} + \Sigma_{xc}(\omega)-U_{xc}\,.
\label{eq:dFdGdetail}
\end{equation}
Setting this matrix derivative to zero locates an saddle and automatically yields the Dyson equation (\ref{eq:dysoneq}) for the Green's function.

\begin{figure}[t!]
\includegraphics[width=3in]{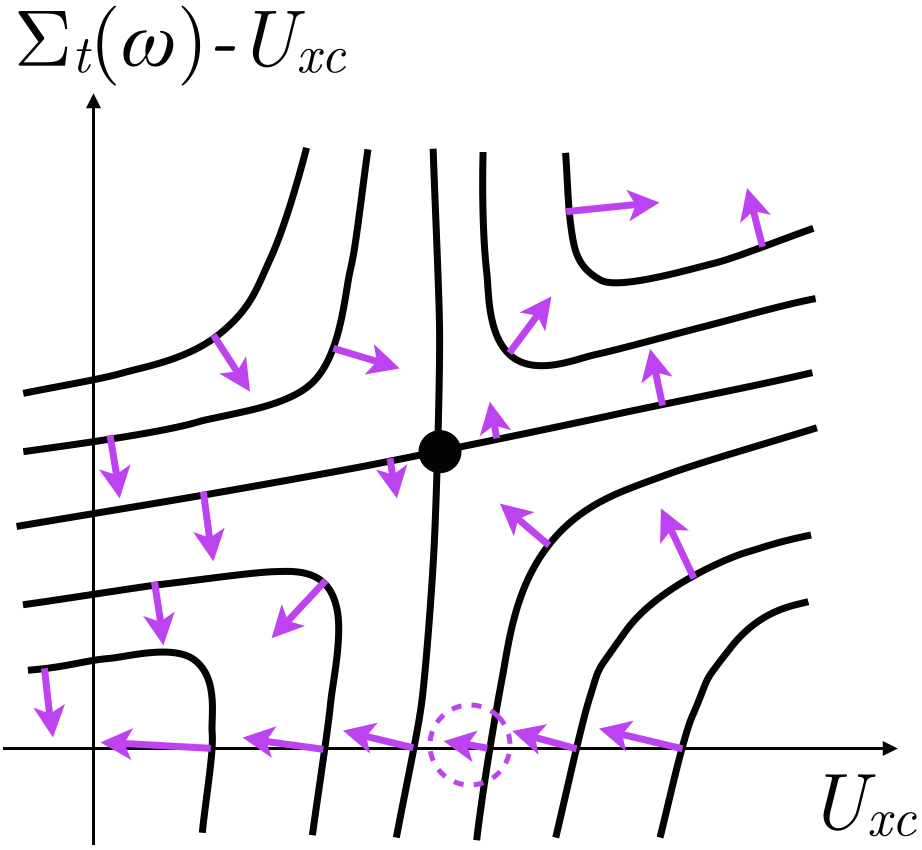}
\caption{Schematic showing the stationary point of $F[G,G_0]$ (black circle), level curves of $F$ (solid lines), and the gradient vector $D(\omega)$ from equation (\ref{eq:domega}) (purple arrows).  The axes represent the choices of trial self-energies $\Sigma_t(\omega)$ in Eq.~(\ref{eq:trialsigma}) that parameterize the Green's function $G$.  The saddle point corresponds to choosing $\Sigma_t$ to be the self-energy $\Sigma_{xc}$ that generates the true Green's function {\it via} the Dyson equation. The gradients evaluated along the horizontal axis $U_{xc}$, representing the subspace of non-interacting Green's functions, are $D_0(\omega)$ of Eq.~(\ref{eq:d0omega}) which are the lowest row of arrows (pointing mainly leftwards).  The objective is to find the $U_{xc}$ that gives the shortest gradient along the $U_{xc}$ axis which in this example is indicated by the dashed circle.
}
\label{fig:contourswithgradients}
\end{figure}

In what follows, it will be convenient to change variables.  Instead of varying $G$ directly, we will vary $G$ via variation of a trial self-energy $\Sigma_t$:
\begin{equation}
G^{-1}(\omega) = G_0^{-1}(\omega) - \left[ \Sigma_t(\omega) - U_{xc}\right]
\label{eq:trialsigma}
\end{equation}
Choosing $\Sigma_t$ to coincide with the  self-energy $\Sigma_{xc}$ that solves the Dyson equation  locates the saddle point of $F$ as per Eq.~(\ref{eq:dFdG}).  Matrix differentiation of $G$ gives
\[
\delta G(\omega) = G(\omega)\delta\Sigma_{t}(\omega)G(\omega)\,.
\]
Using the cyclicity of the trace, the variation of $F$ is 
\begin{equation}
\delta F  =  Tr \Big\{G\left[G^{-1}-G_0^{-1} + \Sigma_{xc}-U_{xc} \right]G \, \delta \Sigma_{t}\Big\}
\label{eq:dFdSigxc}
\end{equation}
which corresponds to the matrix derivative
\begin{equation}
D(\omega)\equiv 2\pi i\frac{\delta F}{\delta \Sigma_{t}(\omega)} = G\Big(G^{-1}-G_0^{-1} + \Sigma_{xc}-U_{xc}\Big)G
\label{eq:domega}\,.
\end{equation}
 We are interested in the case when $G$ is non-interacting, so we set $G=G_0$ and arrive at the simpler derivative
\begin{equation}
D_0(\omega)\equiv D(\omega)\Big|_{G=G_0} = G_0\Big(\Sigma_{xc}(\omega)-U_{xc}\Big)G_0\,.
\label{eq:d0omega}
\end{equation}
Our objective will be to minimize the squared length of the matrix $D_0(\omega)$ 
\begin{equation*}
||D_0||^2  \equiv  \int_{-\infty}^\infty d\omega \, tr\{ D_0(\omega)^\dag D_0(\omega)\} \\
\end{equation*}
and thereby find a gradient-optimal $U_{xc}$ and associated $G_0$.  The situation is shown schematically in Figure~\ref{fig:contourswithgradients}.  Among the set of non-interacting Green's functions parameterized by $U_{xc}$, we are searching for the $U_{xc}$ that makes the gradient $D_0(\omega)$ shortest.  Figuratively, we have constrained ourselves to be along the horizontal axis of the Figure, and we scan along that axis to find the shortest gradient.  

We note that we are not seeking the shortest gradient vector projected into the subspace of non-interacting Green's functions.  That would correspond to examining how $F$ changes due to variations $\delta U_{xc}$  which is different from the much larger set of self-energy variations $\delta \Sigma_{t}(\omega)$.  We are   varying the Green's function both along and away from the non-interacting axis and thus in any arbitrary direction.  This is indicated in Fig.~\ref{fig:contourswithgradients} by the fact that the arrows representing the gradient have components both along and perpendicular to the horizontal axis.

We  aim to minimize $|\delta F/\delta \Sigma_t|$ which is the length of the gradient of $F$ versus the trial self-energy $\Sigma_t$.  As we will see below, it is generally impossible to make the length zero when restricting ourselves to non-interacting Green's functions $G_0$.  Hence, the choice of variable for the derivative will, in principle, change the resulting optimum.  For example, one could try to minimize $|\delta F/\delta G|$ from Eq.~(\ref{eq:dFdGdetail}) instead, and one would arrive at a different set of conditions.  Therefore, choosing the variable for the derivative requires some physical motivation.  Given the choice between $G(\omega)$ and $\Sigma_t(\omega)$, $\Sigma_t$ is a much better choice: choosing $G$ gives useless or nonsense results as detailed in  \ref{appc}.

To progress from these  graphical ideas to analytic formulae, we calculate the squared norm $||D_0||^2$ in the basis of orthonormal eigenstates $\ket{n}$ by inserting complete sets of states:
\begin{equation*}
||D_0||^2 = \int_{-\infty}^\infty d\omega \, \sum_{n,m} |\bra{n}D_0(\omega)\ket{m}|^2\,.
\end{equation*}
Since $G_0$ is diagonal in this basis, this turns into
\begin{multline}
||D_0||^2  =  \int_{-\infty}^\infty \!\! d\omega \sum_{n,m} \frac{|\bra{n}\Sigma_{xc}(\omega)-U_{xc}\ket{m}|^2}
{|\omega-\epsilon_n\pm i\gamma|^2|\omega-\epsilon_m\pm i\gamma|^2}\\
 = \int_{-\infty}^\infty \!\! d\omega \sum_{n,m} \frac{|\bra{n}\Sigma_{xc}(\omega)-U_{xc}\ket{m}|^2}
{[(\omega-\epsilon_n)^2+\gamma^2][(\omega-\epsilon_m)^2+\gamma^2]}\,.
\label{eq:d02formula}
\end{multline}
As expected, this is a sum of strictly positive terms.  Our aim is to choose a $U_{xc}$ so that   $||D_0||^2$ is as small as possible.  

We will be using contour integration techniques to evaluate the frequency integral in Eq.~(\ref{eq:d02formula}).  To do this, we need to examine the self-energy $\Sigma_{xc}(\omega)$ in  more detail.  Most generally, the self-energy along the real $\omega$ axis can be written as a static term plus a sum over poles
\begin{equation}
\Sigma_{xc}(\omega) = \Sigma_x + \sum_a \frac{\sigma_a}{\omega-\xi_a} 
\label{eq:sigxcpoles}
\end{equation}
where $\Sigma_x$ is the static bare exchange (Fock) operator
\[
\Sigma_x(x,x') = -\rho(x,x')/||\vec r - \vec r\,'||\,.
\]
The energies $\xi_a$ locate the poles of the self-energy which have residues given by the matrices $\sigma_a$.  Physically, a pole of $\Sigma_{xc}(\omega)$ occurs at an energy where there is strong quasiparticle scattering by electronic excitations, strongly reduced quasiparticle lifetimes, and a strongly damped  spectral function.  For example, within the $GW$ approximation, these poles correspond to charge fluctuation excitations such as single or multiple electron-hole pairs and plasmons~\cite{Pines,Hedin}.  For a finite system, such as a molecule, the energies $\xi_a$ and associated index $a$ are a discrete and countable set (below the ionization threshold).  Above the ionization threshold or in a solid-state system, there are continuous energy bands and thus a continuum of excitations so the sum over $a$ will be an integral and the self-energy will have branch cuts as a function of $\omega$.  We note that it is possible to have discrete (bound) states embedded in continuum of states~\cite{hsu_bound_2016}, and we will return to these distinctions below when comparing different contributions to the integral of Eq.~(\ref{eq:d02formula}).  

Since $\Sigma_{xc}(\omega)$ \red{either remains finite or grows sublinearly with $\omega$} 
as $|\omega|\rightarrow\infty$ (see \ref{appd}), and the denominator of Eq.~(\ref{eq:d02formula}) grows as $\omega^4$, we can safely turn the integral in Eq.~(\ref{eq:d02formula}) into a closed contour integral which we choose to go over a half circle at infinity in the upper half plane (the lower half plane gives the same results).  Therefore, the quantity we aim to study and minimize is 
\begin{equation}
||D_0||^2  = \sum_{n,m} S_{nm}
\label{eq:d02contour}
\end{equation}
where
\begin{equation}
S_{nm} \equiv \oint d\omega \frac{|\bra{n}\Sigma_{xc}(\omega)-U_{xc}\ket{m}|^2}
{[(\omega-\epsilon_n)^2+\gamma^2][(\omega-\epsilon_m)^2+\gamma^2]}\,.
\label{eq:snmdef}
\end{equation}
We now perform the contour integral separately for diagonal and off-diagonal contributions to Eq.~(\ref{eq:d02contour}) since they will scale differently as a function of $\gamma$.  As a technical aside, we keep in mind that the numerator $|\bra{n}\Sigma_{xc}(\omega)-U_{xc}\ket{m}|^2$ in Eq.~(\ref{eq:d02contour}) begins as a function defined along the real $\omega$ axis: when extending it to complex $\omega$, we simply substitute in  complex $\omega$ into the original functional form ({\it i.e.}, it is only a function of $\omega$ and not of $\omega^*$).  

\subsection{Diagonal terms}

Consider a diagonal contribution $S_{nn}$ to Eq.~(\ref{eq:d02contour})
\begin{equation}
S_{nn} = \oint d\omega\frac{|\bra{n}\Sigma_{xc}(\omega)-U_{xc}\ket{n}|^2}
{[(\omega-\epsilon_n)^2+\gamma^2]^2}\,.
\label{eq:snndef}
\end{equation}
There will be two distinct classes of poles contributing to $S_{nn}$.  

The first and obvious one is the pole coming from the zero of the denominator  at $\omega=\epsilon_n+ i\gamma$ in the upper half plane.  More precisely, we have
\[
[(\omega-\epsilon_n)^2+\gamma^2]^2
 = [\omega-\epsilon_n-i\gamma]^2[\omega-\epsilon_n+i\gamma]^2
\]
and by using the standard Cauchy integral formula 
\[
f'(a) = \frac{1}{2\pi i}\oint \frac{f(z)dz}{(z-a)^2}
\]
the contribution from the pole at $\epsilon_n+i\gamma$ is
\begin{multline*}
2\pi i \frac{d}{d\omega}\left\{
\frac{|\bra{n}\Sigma_{xc}(\omega)-U_{xc}\ket{n}|^2}
{[\omega-\epsilon_n+i\gamma]^2}
\right\}\Big|_{\omega=\epsilon_n+i\gamma}  = \\
 \frac{\pi}{2\gamma^3}|\bra{n}\Sigma_{xc}(\omega)-U_{xc}\ket{n}|^2\Big|_{\omega=\epsilon_n+i\gamma}\\
-\frac{\pi i}{2\gamma^2}\frac{d}{d\omega} \left\{ |\bra{n}\Sigma_{xc}(\omega)-U_{xc}\ket{n}|^2\right\}\Big|_{\omega=\epsilon_n+i\gamma}\,.
\end{multline*}
We now series expand in the small parameter $\gamma$, and find that the imaginary terms scaling as $\gamma^{-2}$ cancel as they must since the integral is manifestly positive and real-valued.  We end  with
\[
 \frac{\pi}{2\gamma^3}|\bra{n}\Sigma_{xc}(\epsilon_n)-U_{xc}\ket{n}|^2 + O(\gamma^{-1})
\]
and note that the dominant contribution scales as $\gamma^{-3}$.

The second set of contributions to $S_{nn}$ will be from poles of $\Sigma_{xc}(\omega)-U_{xc}$ located above the real axis, {\it i.e.}, those with $Im\ \xi_a>0$.  The analysis of the contributions from these poles is straightforward but somewhat long-winded and is detailed in  \ref{appa}.  The result is that the contributions from these poles are subleading:  for systems containing discrete energy levels the contributions scale as $\gamma^{-1}$ while for systems with only continuous energy bands they  scale as $\gamma^0$.

All together, we have 
\[
S_{nn} = A\gamma^{-3}  + O(\gamma^{-1})
\]
where  $A$ originates from the pole at $\epsilon_n+i\gamma$ and is proportional to
\[
A\propto |\bra{n}\Sigma_{xc}(\epsilon_n)-U_{xc}\ket{n}|^2 \,.
\]
We  focus on reducing the magnitude of the large ratio $\bra{n}\Sigma_{xc}(\epsilon_n)-U_{xc}\ket{n}/\gamma$ to make $S_{nn}$ as small as possible since it is dominated by the leading $A\gamma^{-3}$ term.

  In the most general case, the self-energy $\Sigma_{xc}$ will have both real and imaginary parts.  By our assumption of time-reversal symmetry, $U_{xc}$ and the eigenstates of $H_0$ are real-valued.  So we split the matrix element into real and imaginary parts
\[
\bra{n}\Sigma_{xc}(\epsilon_n)-U_{xc}\ket{n} = R + i I
\]
where
\[
R = \bra{n}Re\ \Sigma_{xc}(\epsilon_n)-U_{xc}\ket{n}
\]
and
\[
I = \bra{n}Im\ \Sigma_{xc}(\epsilon_n)\ket{n}\,.
\]
Thus the leading term in $S_{nn}$ is
\[
A\gamma^{-3} = \frac{\pi}{2\gamma^3}\left[ R^2 + I^2\right]\,.
\]
Mathematically, it is  unlikely that one can set $A=0$ since the only free variable is the single real-valued $\bra{n}U_{xc}\ket{n}$ while there are two functions $R$ and $I$ to set to zero.  Physically, this corresponds to the fact that $U_{xc}$ is Hermitian so it can adjust real-valued energies but never give an imaginary part to the energy which would correspond to a quasiparticle lifetime; such lifetimes are described by non-zero $Im\ \Sigma_{xc}$.  Thus we can  set $R=0$ by choosing $\bra{n}U_{xc}\ket{n} = \bra{n}\Sigma_{xc}(\epsilon_n)\ket{n}$ while we must settle for a non-zero $I$ in the general case.

Luckily, there are many cases where either $I=0$ or it is not the main quantity of physical consideration.  In systems with discrete  energy eigenstates, the lifetimes of excitations will be infinite and thus $I=0$.  In solid state systems with continuous energy bands that have an energy gap, quasiparticles whose energies are within one energy gap of either the valence or condition band edges also have infinite lifetimes due to electron-electron interactions  since there are no lower-energy states for them to decay into while conserving overall energy.  Finally, in most practical $GW$ calculations, one  focuses on the real part of $\Sigma_{xc}$ in order to correct DFT band energies ({\it e.g.}, Refs.~\cite{HL,scCOHSEX,QPSCGW}).  In all these cases, when we choose $\bra{n}U_{xc}\ket{n} = \bra{n}\Sigma_{xc}(\epsilon_n)\ket{n}$,  $A$ becomes zero and this  reduces the scaling of $S_{nn}$ from $O(\gamma^{-3}$) to $O(\gamma^{-1}$).

Summarizing this subsection, for small $\gamma$, choosing the matrix element of $U_{xc}$ to obey $\bra{n}U_{xc}\ket{n} = \bra{n}\Sigma_{xc}(\epsilon_n)\ket{n}$ is the choice that will make $S_{nn}$ as small as possible.  For cases where the imaginary part of $\Sigma_{xc}$ is zero because (i) one has discrete energy spectra, (ii) the system has an energy gap and one is focused on states near the band edges, or (iii) one is ignoring the imaginary part because one is focused on the real part of $\Sigma_{xc}$ in order to predict band energies, this choice leads to a significant reduction of $S_{nn}$ from scaling as $\gamma^{-3}$ to scaling as $\gamma^{-1}$.  In situations where we also include $Im\ \Sigma_{xc}$, this choice  is  the most sensible and obvious.  However, the reduction is more modest: the coefficient of the $\gamma^{-3}$ scaling is reduced and becomes$~\sim|\bra{n}Im\ \Sigma_{xc}(\epsilon_n)\ket{n}|^2$.  Either way, this choice requires a self-consistent process because the energy $\epsilon_n$ and the self-energy $\Sigma_{xc}$ both depend on $U_{xc}$.

\subsection{Off-diagonal terms}

For a general off-diagonal contribution with $n\ne m$
\begin{equation}
S_{nm} = \oint d\omega \frac{|\bra{n}\Sigma_{xc}(\omega)-U_{xc}\ket{m}|^2}
{[(\omega-\epsilon_n)^2+\gamma^2][(\omega-\epsilon_m)^2+\gamma^2]}
\label{eq:Snmoffdiag}
\end{equation}
we have two simple poles above the real $\omega$ axis at $\epsilon_n+i\gamma$ and $\epsilon_m+i\gamma$ as well as the poles of the numerator stemming from $\Sigma_{xc}(\omega)$.  The two simple poles contribute the following term
\begin{equation}
\frac{\pi}{\gamma}\!\cdot\!\frac{|\bra{n}\Sigma_{xc}(\epsilon_n)\!-\!U_{xc}\ket{m}|^2 \!+\! |\bra{n}\Sigma_{xc}(\epsilon_m)\!-\!U_{xc}\ket{m}|^2}{(\epsilon_n-\epsilon_m)^2}\!\!
\label{eq:nmpolescontrib}
\end{equation}
which scales as $\gamma^{-1}$.  For the moment, we ignore the additional contributions coming from the poles of $\Sigma_{xc}(\omega)$ and instead focus on minimizing the above contribution from the two simple poles.  Specifically, our objective is to choose the $U_{xc}$ that minimizes the above expression.  We envisage this as a self-consistent process where (i) we hold all quantities fixed except for $\bra{n}U_{xc}\ket{m}$ which is allowed to vary to optimize the expression,  (ii) we update all quantities using the new $U_{xc}$, and (iii) we iterate to convergence.

To simplify the algebra, we define $z=\bra{n}U_{xc}\ket{m}$, $\kappa_n = \bra{n}\Sigma_{xc}(\epsilon_n)\ket{m}$, and $\kappa_m = \bra{n}\Sigma_{xc}(\epsilon_m)\ket{m}$.  Keeping in mind that $z$ is real-valued due to our assumption of time-reveal invariance, the expression to be optimized is quadratic in $z$:
\[
\frac{\pi}{\gamma}\cdot \frac{
|\kappa_n|^2+|\kappa_m|^2 - 2z Re(\kappa_n+\kappa_m) +2 z^2
}{(\epsilon_n-\epsilon_m)^2}
\]
Setting the derivative of this quadratic versus $z$ to zero, we find the optimum
\[
z = Re(\kappa_n+\kappa_m)/2
\]
or in other words
\[
\bra{n}U_{xc}\ket{m} = Re\left\{\frac{\bra{n}\Sigma_{xc}(\epsilon_n)\ket{m} + \bra{n}\Sigma_{xc}(\epsilon_m)\ket{m}}{2}\right\}\,.
\]

This choice is guaranteed to minimize the contributions to $S_{nm}$ scaling as $\gamma^{-1}$ that originate from the simple poles at $\epsilon_n+i\gamma$ and $\epsilon_m+i\gamma$.  For a system with only continuous energy bands, this is a good choice since the contributions coming from the numerator ({\it i.e.}, the poles of $\Sigma_{xc}$) are subleading and scale as $\gamma^0$ as shown in  \ref{appa}.  However, for a system containing discrete energy levels, the contributions from the poles of the numerator also scale as $\gamma^{-1}$ so that the above considerations do not provide an air-tight argument.  In \ref{appe}, expressions for these contributions and estimates of their size are provided in the context of optimizing the gradient once self-consistency (QS) is achieved. Whether one should be ignoring them or one must include them in the optimization  is a subject for further investigation.  Finally, as we saw in the case of the diagonal contribution $S_{nn}$, the optimal Hermitian $U_{xc}$ is only determined by the real part of $\Sigma_{xc}$.

\subsubsection{Off-diagonal case with degeneracy}

The above discussion of the off-diagonal case assumed that $\epsilon_n \ne \epsilon_m$.  Specifically, in going from the contour integral of Eq.~(\ref{eq:Snmoffdiag}) to the result of Eq.~(\ref{eq:nmpolescontrib}) it was assumed that $|\epsilon_n-\epsilon_m|>\gamma$ so that we had two distinct poles contributing.  The correct way of proceeding in the case of degeneracy $\epsilon_n=\epsilon_m$ is to return to Eq.~(\ref{eq:Snmoffdiag}) and notice that the denominator becomes identical in structure to that of the diagonal case in Eq.~(\ref{eq:snndef}).  In fact, the only difference with the diagonal case is the replacement of $\bra{n}\Sigma_{xc}(\omega)-U_{xc}\ket{n}$ by $\bra{n}\Sigma_{xc}(\omega)-U_{xc}\ket{m}$ while the remainder of the analysis remains identical.   We end up with the optimal choice $\bra{n}U_{xc}\ket{m}=Re\bra{n}\Sigma_{xc}(\epsilon_n)\ket{m}$ in this degenerate case.  It is gratifying that this is identical to the optimal $U_{xc}$ in the non-degenerate off-diagonal case where we simply set $\epsilon_n=\epsilon_m$.

\subsection{Discussion}

The main result is that the length of the gradient of the Klein energy functional $F$ is minimized when $U_{xc}$ is chosen to satisfy
\begin{equation}
\bra{n}U_{xc}\ket{m}\! =\! Re\!\left\{\!\frac{\bra{n}\Sigma_{xc}(\epsilon_n)\ket{m}\! + \! \bra{n}\Sigma_{xc}(\epsilon_m)\ket{m}}{2}\!\right\}\!\!
\label{eq:qsgw}
\end{equation}
when $\gamma$ is small.   

This choice of $U_{xc}$ is identical to that of the QS$GW$ method.  In QS$GW$, one approximates the self-energy to its $GW$ form {\it and} one sets the imaginary part of the self-energy to zero.  The QS$GW$ has successfully described the band structure of a wide variety of solid state systems within the $GW$ approximation for the self-energy~\cite{QPSCGW}.  Therefore, in addition to its practical successes, we can say that the $QS$GW is also mathematically well-founded as it is the choice for $U_{xc}$ that minimizes the length of the gradient of the energy functional when approximated within $GW$. It is ``closest'' to the interacting $G$ obeying Dyson's equation. 

A critical  point of the above derivation is that it is not dependent on the $GW$ approximation itself:  the optimum choice of Eq.~(\ref{eq:qsgw}) holds for {\it any} self-energy $\Sigma_{xc}(\omega)$ at any level of approximation as long as it is derived from some $\Phi_{xc}[G]$ {\it via} $\Sigma_{xc}=2\pi i\delta \Phi_{xc}/\delta G$.  Namely, if we assume that the shortest gradient of the energy functional is best, the recipe of Eq.~(\ref{eq:qsgw}) is a general answer to the problem of choosing the best non-interacting Green's function to describe an interacting system.

We also note the significant difference between diagonal and off-diagonal elements.  Having the diagonal elements obey Eq.~(\ref{eq:qsgw}) reduces the length of the gradient by factors of $O(\gamma^{-3})$.  If the imaginary part of the self-energy is zero or set to zero, then the reduction is very strong: the diagonal contributions in fact become reduced to $O(\gamma^{-1})$.  On the other hand, obeying Eq.~(\ref{eq:qsgw}) for the off-diagonal elements doesn't actually change the scaling --- off-diagonals always contribute $O(\gamma^{-1}$) to the length of the gradient --- but it does reduce the magnitude of the coefficients of the terms scaling as $\gamma^{-1}$.  Therefore, from a practical point of view, obeying Eq.~(\ref{eq:qsgw}) for the diagonal elements is of primary importance while obeying it for off-diagonals is of secondary importance.  This is a way of rationalizing the observation, dating back to the earliest fully {\it ab initio} GW calculations~\cite{HL}, that in many (but not all) cases the most critical corrections to the quasiparticle properties are handled by the diagonal terms of the self-energy.

\section{Smallest energy change $\Delta F$}
\label{sec:deltaF}

\begin{figure}[t!]
\includegraphics[width=3in]{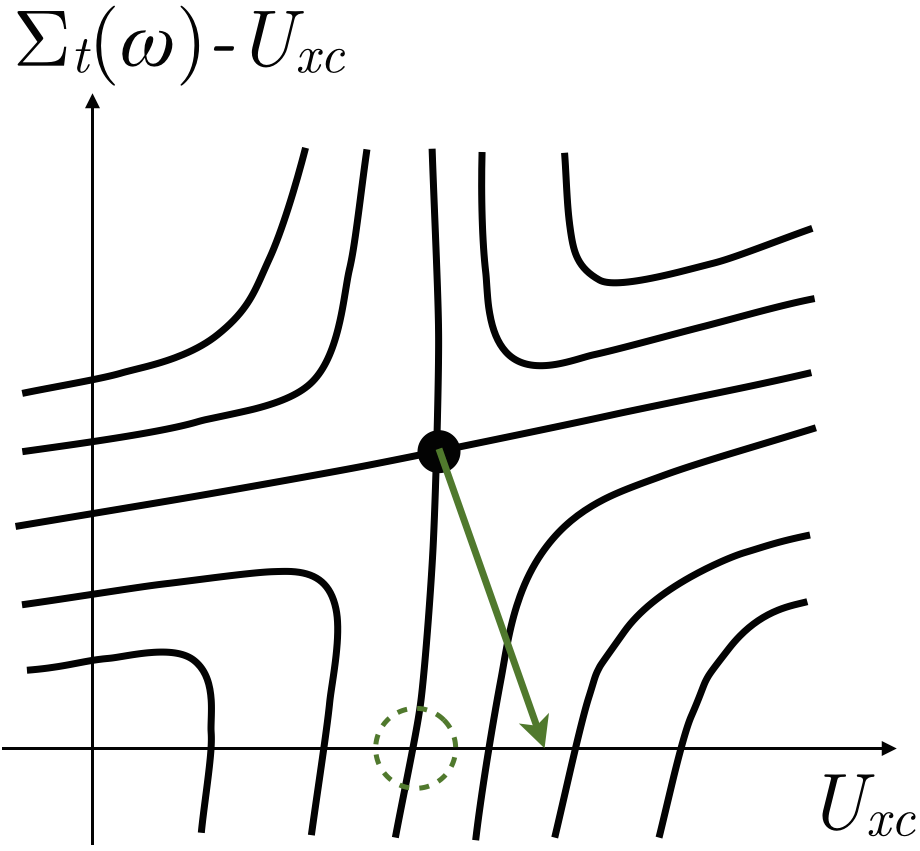}
\caption{Please see the caption of Fig.~\ref{fig:contourswithgradients} for meaning of axes, {\it etc.}. The green vector $U_{xc}-\Sigma_{xc}$ represents the deviation of $U_{xc}$ from the stationary self-energy $\Sigma_{xc}$ and connects the saddle point to the chosen $U_{xc}$ on the horizontal axis.  The dashed circle represents the most desirable $U_{xc}$ since at that point the energy $F$ has the same value as it does at the saddle point.}
\label{fig:vectortouxc}
\end{figure}

A alternative approach to quantifying which non-interacting Green's function $G_0$ is ``best'' is to try to find the $G_0$ that generates the smallest deviation of $F$ from its value at the saddle point.  Namely, when scanning along the horizontal axis of Fig.~\ref{fig:contours}, one looks for the $U_{xc}$ that generates the $G_0$ so that $F[G_0,G_0]$ is as close as possible to the true total energy $F[\bar G,G_0]$ where $\bar G$ solves the Dyson Eq.~(\ref{eq:dysoneq}).  Specifically, what we would like to minimize is the magnitude of the energy difference $\Delta F$ defined as
\begin{equation}
\Delta F[G_0] \equiv F[G_0,G_0] - F[\bar G,G_0]\,.
\end{equation}
To make headway analytically, we will assume that the ``best'' $G_0$ is sufficiently close to the saddle point so that the difference $\bar G-G_0$ or equivalently $\Sigma_{xc}-U_{xc}$ is small enough for a quadratic approximation of  $F$ to be accurate.  With the quadratic assumption, we can use the general fact that the value of a quadratic function is given by half the dot product of its gradient times the displacement from its stationary point.  From Eq.~(\ref{eq:dFdSigxc}), the gradient of $F$ is $G\left[G^{-1}-G_0^{-1} + \Sigma_{xc}-U_{xc} \right]G$.  The  displacement is $U_{xc}-\Sigma_{xc}$.  We are evaluating all these expressions at $G=G_0$.  The situation is graphically illustrated in Fig.~\ref{fig:vectortouxc}.
Therefore, the quadratic approximation form for $\Delta F$ is~\cite{higherordernote}
\begin{equation}
\Delta F = -\frac{1}{2}Tr\left\{
G_0\left[\Sigma_{xc}-U_{xc} \right]G_0 ( \Sigma_{xc}-U_{xc})
\right\}\,.
\label{eq:deltaF}
\end{equation}
An explicit expression in terms of integrals and matrix elements is
\begin{equation}
\Delta F = -\frac{1}{2}
\int_{-\infty}^\infty  \frac{d\omega}{2\pi i}e^{i\omega 0^+} \sum_{n,m}\\
\frac{\bra{n}\Sigma_{xc}(\omega)-U_{xc}\ket{m} \bra{m}\Sigma_{xc}(\omega)-U_{xc}\ket{n}}{(\omega-\epsilon_n-i\gamma s_n)(\omega-\epsilon_m-i\gamma s_m)}\,.
\label{eq:deltaFnm}
\end{equation}

One can proceed by closing the integrals over the upper complex half plane and computing the residues of the integral with separate contributions from the poles in the denominator as well as the poles of $\Sigma_{xc}(\omega)$ in the numerator.  \ref{appb} contains the details which produce  algebraic expressions that do not --- for this author ---  provide  insight into how one should proceed.

Aside from the algebraic complexities, there are two other higher level challenges with this approach.  First, one is trying to reduce the magnitude of $\Delta F$ or equivalently make it as close to zero as possible.   However, since we are close to a saddle point, $\Delta F$ will take both positive and negative values which makes the optimization much more challenging than the minimization of a function bounded from below.  Second, unlike the previous approach of minimizing the length of the gradient, there is no obvious small parameter such as $\gamma$ to permit us to perform the optimization process in an organized fashion and to identify the largest terms that must be handled first.  Hence, either this smallest-$\Delta F$-approach is inherently difficult, or a new idea is needed that will take it in a more successful direction.  This is an open question.

\section{Application of shortest gradient}
\label{sec:application}

In this section, we apply the shortest gradient approach and its associated quasiparticle self-consistent (QS) scheme to understand its behavior in interacting electronic systems.  Since this approach justifies practical schemes such as the QSGW~\cite{QPSCGW}, the success of QSGW in predicting realistic electronic properties of a wide class of materials may be viewed as an ``application'' of the method.  Of course, at its best, QSGW is only as accurate as the underlying $GW$ approximation for the self-energy.

We apply the shortest gradient approach to a solvable model many-body system for which we know the exact self-energy.  In this way, we can make some conclusions about how this approach and its associated QS scheme describe a many-body system.  We focus on a situation where there are multiple quasiparticle solutions corresponding to a single non-interacting state.  The existence of multiple solutions is a hallmark of an interacting many-body system, typically with a multi-determinant ground state.  A number of realizations of this multiplicity are found in the literature for model systems as well as realistic molecules and materials~\cite{pavlyukh_configuration_2007,lischner_first-principles_2012,pavlyukh_vertex_2016,caruso_gw_2016}.

We study a two-site Hubbard model with single orbital per site and single electron per site: this Hamiltonian has also been found useful in studying multiple solution situations in prior work~\cite{lischner_first-principles_2012}.  The Hamiltonian is
\[
\hat H = -t \sum_{\sigma\in\{\uparrow,\downarrow\}}\left( \hat c_{1,\sigma}^\dag 
\hat c_{2,\sigma} + \hat c_{1,\sigma}^\dag \hat c_{2,\sigma} \right)
+ U \sum_{i=1,2} \hat n_{i,\uparrow}\hat n_{i,\downarrow}
\]
where $\hat n_{i\sigma} = \hat c_{i,\sigma}^\dag \hat c_{i,\sigma}$, $\sigma$ is the spin index, and $i\in\{1,2\}$ labels the two sites.  Due to its simplicity, high symmetry, and small Hilbert space, this Hamiltonian is readily diagonalized by hand for any number of  electrons $N\in\{0,1,2,3,4\}$.  The  ground state for $N=2$ is a spin singlet.
With some minor effort, one can compute the exact one particle Green's functions analytically.  

As expected from symmetry considerations and the singlet ground state, the Green's function is diagonal in spin and also diagonal in the basis of bonding ($b$) and anti-bonding ($a$) single-particle states,
\[
\ket{b,\sigma} = \frac{1}{\sqrt{2}}\left( \ket{1,\sigma} + \ket{2,\sigma}\right) \ \ , \ \ 
\ket{a,\sigma} = \frac{1}{\sqrt{2}}\left( \ket{1,\sigma} - \ket{2,\sigma}\right)\,,
\]
with non-interacting energy eigenvalues $\epsilon^0_b=U/2-t$ and $\epsilon^0_a=U/2+t$ (the constant $U/2$ is the Hartree potential which we include in the non-interacting eigenvalue as per Section~\ref{sec:defs}).
The  quantity of interest to us is the exact self-energy for exchange and correlation which has two separate forms (bonding and anti-bonding):
\begin{equation}
\Sigma_b(\omega) = \frac{U^2/4}{\omega-U/2-3t-i\gamma} \ \ , \ \ \Sigma_a(\omega) = \frac{U^2/4}{\omega-U/2+3t+i\gamma} \,.
\label{eq:simgabaexact}
\end{equation}

\begin{figure}[t!]
\includegraphics[width=3in]{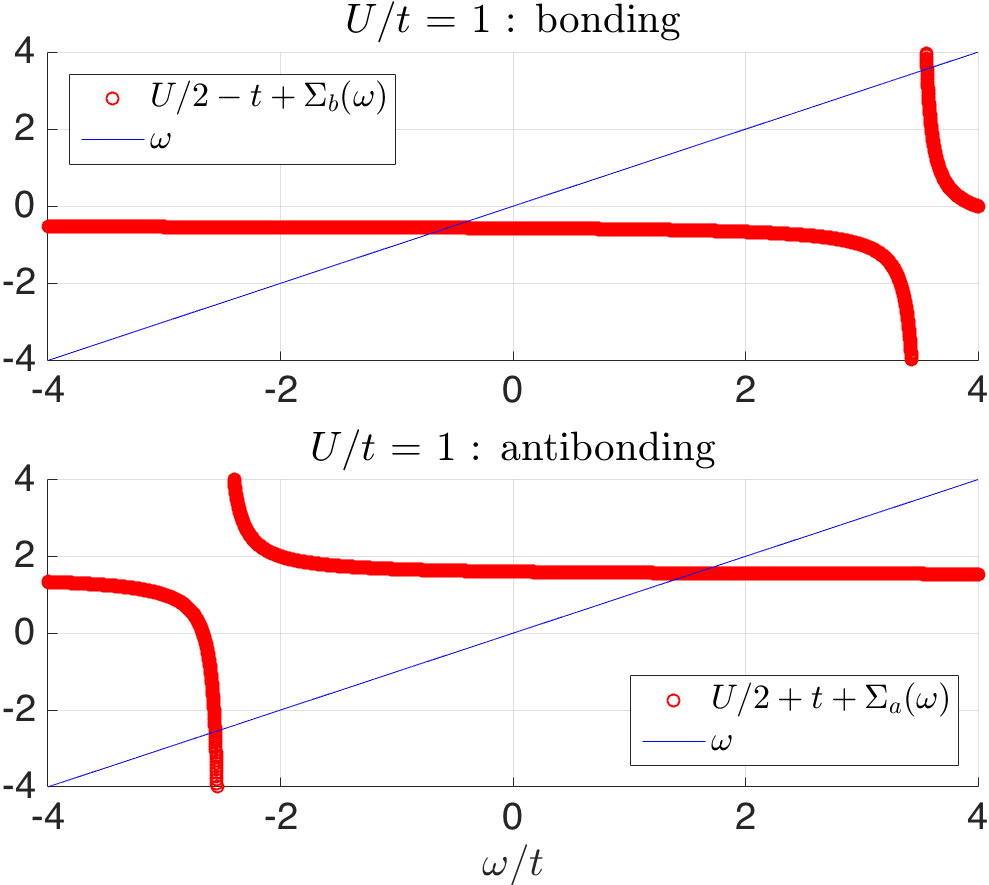}
\includegraphics[width=3in]{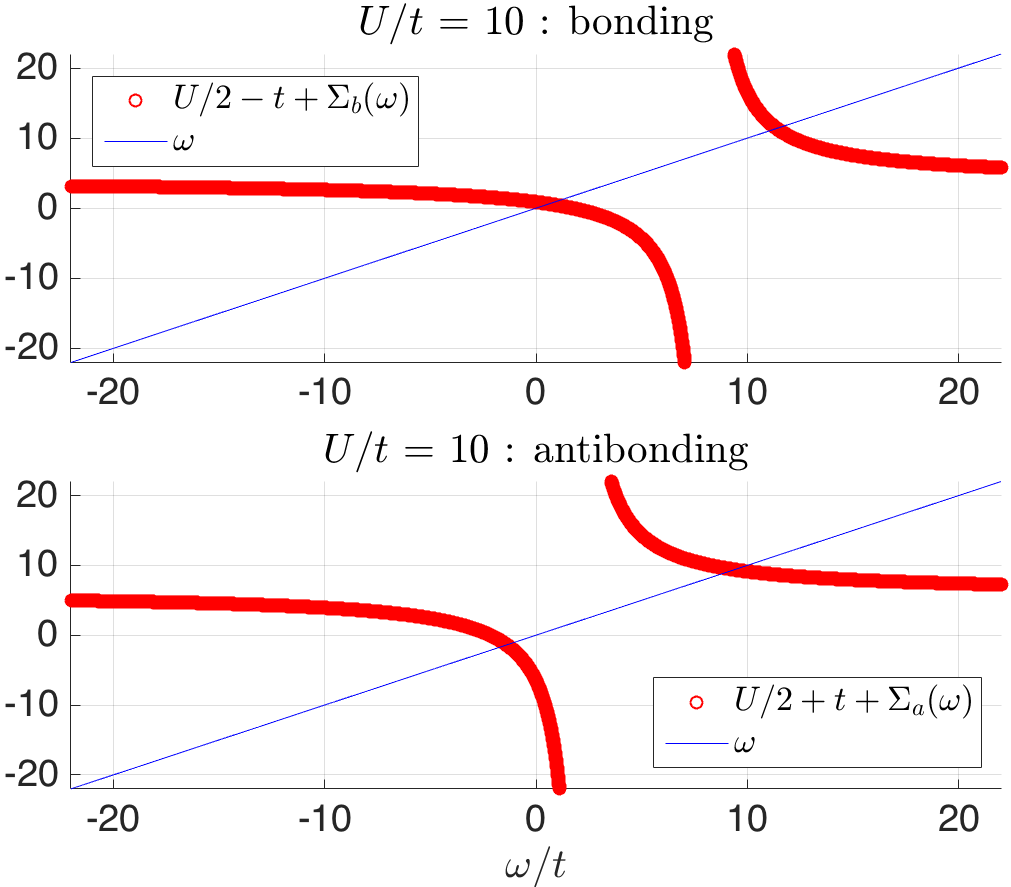}
\caption{Graphical solution to the QS self-consistency condition for the two-site single-orbital Hubbard model at half filling for $U/t=1$ (left) and $U/t=10$ (right).  An energy solution occurs when the thin blue straight line crosses the thick red curve.
}
\label{fig:crossingsUt}
\end{figure}

Due to the diagonal nature of the problem in the $a,b$ basis, the shortest gradient is achieved when the QS condition for the self-consistent single-particle $\epsilon_{i,n}$ is satisfied:
\[
\epsilon_{j,n} - \epsilon^0_j = U_{xc}^{j,n} = Re\Sigma_j(\epsilon_{j,n})\,.
\]
Here $j\in\{a,b\}$, and $n\in\{1,2\}$ labels the two solutions to this equation.  The solutions are never at the poles of $\Sigma_j(\omega)$ so when solving the above equation we set $\gamma=0$. This can be rewritten as the Dyson equation for the eigenvalue:
\[
\epsilon_{j,n} = \epsilon^0_j + Re \Sigma_j(\epsilon_{j,n})\,.
\]
Figure~\ref{fig:crossingsUt} shows the usual graphical solution to this equation for weak interaction $U/t=1$ and strong interaction $U/t=10$.  For each of $a$ and $b$, there are two solutions: one solution occurs where the slope of $\Sigma_j(\omega)$ is large, and the other solution where the slope is small.  Hence, the quasiparticle weights $Z_{j,n}=1/(1-\Sigma_j'(\epsilon_{j,n}))$ for the solutions are small and large, respectively.

\begin{figure}[t!]
\includegraphics[width=4in]{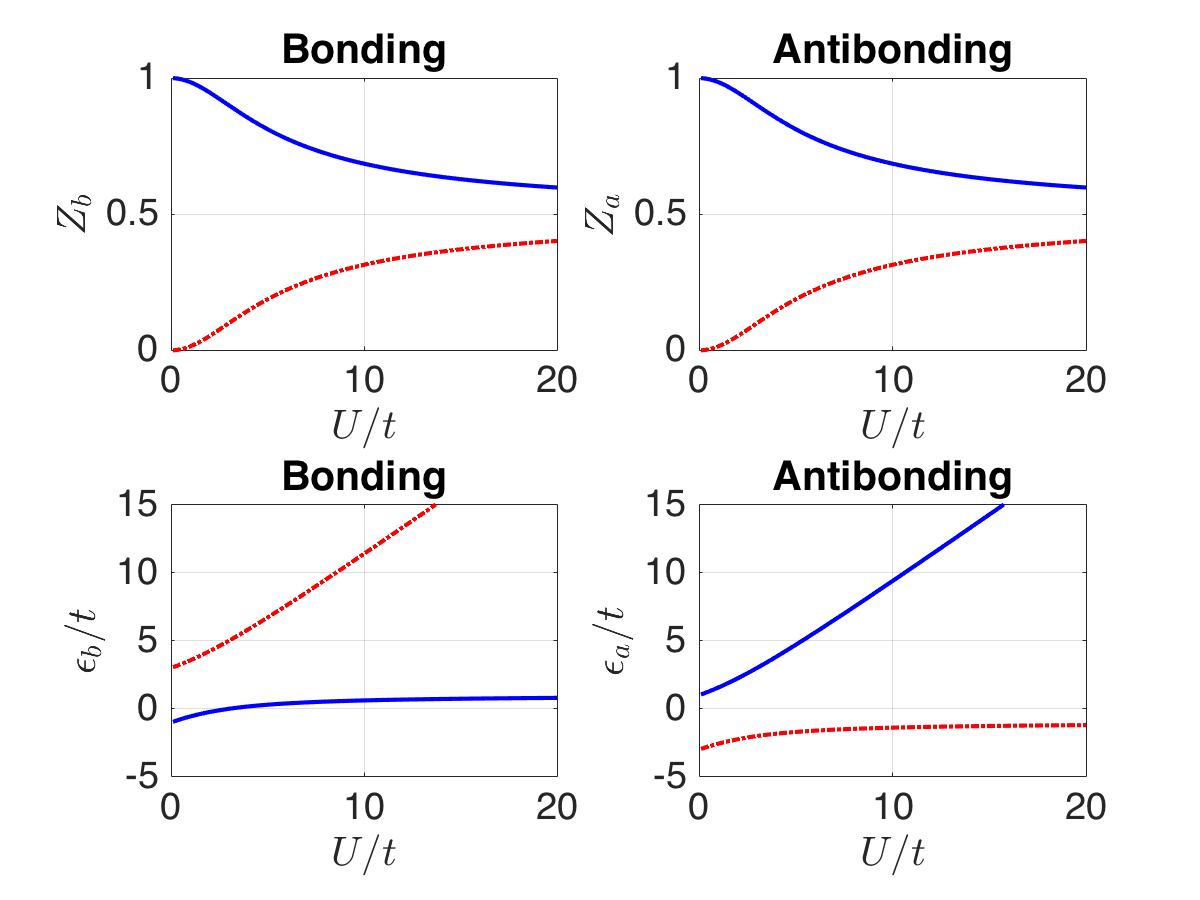}
\caption{Evolution of the QS energy eigenvalues $\epsilon_{j,n}$ and their associated quasiparticle weights $Z_{j,n}$ versus $U/t$ for the two-site single-band Hubbard model at half filling.  Top panels show $Z$ and bottom panels show $\epsilon$, and the colors of the curves correspond for a pair of vertical panels (e.g., the large $Z_b$ bonding solution in solid blue in the upper left corresponds to the lower energy bonding energy $\epsilon_b$ in solid blue in the lower left).
}
\label{fig:ZepsilonvsU}
\end{figure}

Figure~\ref{fig:ZepsilonvsU} shows the evolution of the QS energies and quasiparticle weights as a function of interaction strength $U/t$.  As expected, for small $U/t$, one set of solutions has $Z$ close to unity with energies evolving smoothly into the non-interacting $\epsilon^0_j$  as $U/t\rightarrow0$; the second set of solutions have very small $Z$ in this limit.  For large $U/t$, the system is strongly correlated and does not resemble a single-particle system: all solutions show $Z\rightarrow1/2$ which means that the system becomes impossible to describe with a single Slater determinant.  For $U/t\rightarrow\infty$, the QS energies tend to zero or $U$ which are the ``Hubbard bands'' for this simple system (the lower and upper Hubbard bands each have half the spectral weight in each $a$ or $b$ channel).

Due to the existence of multiple (here two) QS solutions for each non-interacting state ($b$ or $a$), the QS scheme itself does not provide us with a unique solution for the energy states.  For each spin channel, any of the four possible combinations of solutions are self-consistent in the QS sense.  We  now use the criterion of shortest gradient to make a unique choice.  We compute the square length of the gradient from Eq.~(\ref{eq:d02formula}) given by the integral
\[
||D_0||^2 = \sum_{j,\sigma} \int_{-\infty}^\infty d\omega \frac{|\Sigma_j(\omega)-U_{xc}^{j,n}|^2}
{[(\omega-\epsilon_{j,n})^2+\gamma^2]^2}
\]
Due to the high symmetry of the system, in addition to spin degeneracy, the integrals for bonding $j=b$ and anti-bonding $j=a$ have identical values when one uses corresponding solutions (high $Z$ or low $Z$).   Therefore, we  omit the spin index and set $j=b$.  Furthermore, since the integrals will scale as $1/\gamma$ for small $\gamma$, we  focus  on 
\[
I_{b,n} = \gamma  \int_{-\infty}^\infty\frac{|\Sigma_b(\omega)-U_{xc}^{b,n}|^2}
{[(\omega-\epsilon_{b,n})^2+\gamma^2]^2}\,.
\]
We numerically evaluate the integrals  and tabulate the results in Table~\ref{tab:gradtable}.  The table shows that asking for the shortest gradient is a non-ambiguous procedure that clearly favors choosing the large $Z$ solution for any $U/t$.  

An order magnitude estimate of the integral for small $\gamma$ is provided by the linearization $\Sigma_b(\omega)-U^{b,n}_{xc}\approx\Sigma'_b(\epsilon_{b,n})(\omega-\epsilon_{b,n}-i\gamma) $ near the QS energy solution, where the term in $i\gamma$ comes from series expansion of $\Sigma_b(\omega)$ in $\gamma$.  The resulting integrals are ratios of simple polynomials and yield  $\pi |\Sigma'_b(\epsilon_{b,n})|^2 =\pi|1-1/Z_{b,n}|^2$.  However, this underestimates the tabulated value by a factor of two.  This estimate has neglected the contribution from the pole of $\Sigma_{b}(\omega)$ which, as explained in Section~\ref{sec:gradient}, also contributes a term scaling as $1/\gamma$.  In this particular model system, the two contributions are equal.  
\ref{appe} presents   arguments for general cases, beyond this model system, showing that the gradient should be shortest when the solution is chosen that has maximum $Z$ for each non-interacting state.

In this section, we have shown that for systems which present multiple QS  solutions corresponding to a single non-interacting state, gradient optimization will choose the solution with largest quasiparticle weight $Z$.  This is sensible since one is asking for the best non-interacting description of an interacting problem, and states with the largest $Z$ are the most non-interacting (i.e., they are the ones best described by a single Slater determinant ground state).  This is the best we can do with with a non-interacting description: a non-interacting single-particle Hamiltonian $H_0$ will have one eigenvalue for each eigenvector and can not describe multiple solutions for the same vector.  The gradient optimization approach gives a rational basis for choosing solutions with largest $Z$ for each non-interacting state.

\begin{table}
\begin{tabular}{c|cc|cc}
$U/t$ & $Z_{b,1}$ & $I_{b,1}$ & $Z_{b,2}$ & $I_{b,2}$ \\\hline
1 & 0.985 & 1.44$\times10^{-3}$ & 0.0149 & 2.74$\times10^4$\\
4 & 0.854 & 0.185  & 0.146 & 214\\
8 & 0.724 & 0.912 & 0.276 & 43.07\\
12 & 0.658 & 1.70 & 0.342 & 23.3
\end{tabular}
\caption{Scaled squared length of the gradient $I_{b,n}$ for each choice of QS solution associated with quasiparticle weight $Z_{b,n}$.  Values of $I_{b,n}$ listed in the table are found by first numerically integrating for a set of finite and decreasing values of $\gamma$ and then  extrapolating to $\gamma=0$.
}
\label{tab:gradtable}
\end{table}


\section{Summary}
\label{sec:conclusions}

The \red{Baym-Kadanoff} approach provides a total energy functional of trial one-particle Green's functions that has a stationary point at the physically correct Green's function that solves the Dyson equation.  In addition, it provides a recipe for computing the self-energy {\it via} differentiation of an exchange-correlation energy functional.  In practice, dealing with arbitrary Green's functions is computationally complex and also conceptually difficult as the representability criteria for physical Green's functions are not known.  One way forward is to restrict oneself to simpler non-interacting Green's functions.

We have described two approaches to finding the ``best'' non-interacting Green's function.  The first is based on minimizing the magnitude of the derivative of the \red{Klein} energy functional, and this approach produces definite results that form the main body of this paper.  The second approach is based on minimizing the error in total energy of the Klein functional between the trial state and the exact state, but this idea needs further development to be useful.

The gradient minimization approach yields a set of equations for the non-interacting Green's function that are  identical to those of the quasiparticle self-consistent $GW$ (QS$GW$) scheme~\cite{QPSCGW}.  This means that this type of approach has a firm, {\it a priori} foundation.  In addition, we have shown that the resulting quasi-particle self-consistent equations are not unique to the $GW$ approximation but hold for any exchange-correlation functional.  Namely, the equations are the same for any self-energy in \red{Baym-Kadanoff theory when using the Klein functional}.  

Separately, by applying the gradient optimization method to a simple but non-trivial many-body system as well as providing more general arguments, we described how this approach chooses the ``best'' non-interacting Green's function in cases where there are multiple quasiparticle solutions corresponding to the same non-interacting state.  Specifically, gradient optimization favors choosing the solution with largest quasiparticle weight $Z$: this  is  intuitively sensible  when approximating an interacting system with non-interacting single-particle states since the states with largest $Z$ have the largest non-interacting component.

Finally, not only does our work justify quasiparticle self-consistent schemes, but it also provides theoretical insight and justification as to why a ``diagonal-only'' approach for self-energy calculations is a correct starting point and yields good results whereas inclusion of off-diagonal contributions of the self-energy, while physically important in some cases, is of a subdominant nature.   Both findings correlate well with practical experience and observations in the field for {\it ab initio} predictions of electronic properties.

\ack
This work has been supported primarily by the National Science Foundation under Grant No. MRSEC DMR 1119826.

\appendix

\section{Contribution from poles of $\Sigma_{xc}$}
\label{appa}

As stated in the text, when performing the contour integral of Eq.~(\ref{eq:snmdef}), which is reproduced here
\begin{equation}
S_{nm} = \oint d\omega \frac{|\bra{n}\Sigma_{xc}(\omega)-U_{xc}\ket{m}|^2}
{[(\omega-\epsilon_n)^2+\gamma^2][(\omega-\epsilon_m)^2+\gamma^2]}\,,
\label{eq:appendixsnmdef}
\end{equation}
a dominant contribution results from the poles above the real axis associated with the denominator at energies $\epsilon_n+ i\gamma$ and $\epsilon_m+ i\gamma$.  The remaining contributions come from the poles of the numerator $|\bra{n}\Sigma_{xc}(\omega)-U_{xc}\ket{m}|^2$, and in this appendix we examine these poles and the scaling of their contributions to Eq.~(\ref{eq:snmdef}).  We find that the scaling is either $\gamma^{-1}$ or $\gamma^0$ depending on whether the spectrum of poles contains discrete states or only continua (bands), respectively.

Taking the matrix element of the self-energy from Eq.~(\ref{eq:sigxcpoles}) between two states $\ket{n}$ and $\ket{m}$, we have
\beq
\bra{n}\Sigma_{xc}(\omega)-U_{xc}\ket{m} & = & \bra{n}\Sigma_x-U_{xc}\ket{m} + \sum_a \frac{\bra{n}\sigma_a\ket{m}}{\omega-\xi_a} \\
& = & r + \sum_a \frac{s_a}{\omega-\xi_a}
\eeq
where for compactness we have defined $r$ and $s_a$ and  suppressed the $n,m$ indices since we analyze the contribution of a single $(n,m)$ pair.

The pole energies $\xi_a$ can either have a positive imaginary part $Im\ \xi_a \ge \gamma>0$ or a negative one $Im\ \xi_a \le -\gamma<0$. A positive imaginary part represents a process moving backwards in time and thus involves holes, while a negative imaginary part means a forward moving process and thus involves electrons.  The imaginary part can be larger in magnitude than $\gamma$ if the associated excitation is physically damped with a significant decay rate $|Im\ \xi_a|\gg \gamma$, but here we take the worst-case scenario for a conservative  analysis by setting $Im\ \xi_a=\pm\gamma$.

We split the poles $\xi_a$ into the set with positive imaginary part identified with index $b$ and superscript $+$ and the remaining ones with negative imaginary part identified with index $c$ and superscript $-$.  We also separate out the real part of the pole energies $\xi_a^\pm$ and define them to be $\omega_b^+$ and $\omega_c^-$.  Hence, we have
\begin{equation}
\bra{n}\Sigma_{xc}(\omega)-U_{xc}\ket{m}  = r  \\
+\sum_b \frac{s_b^+}{\omega-\omega_b^+-i\gamma} + \sum_c \frac{s^-_c}{\omega - \omega_c^-+i\gamma}
\label{eq:sigpoleform}
\end{equation}

The main physical assumption we will make on the energies $\omega_b^+$ and $\omega_c^-$ is that for any $b$ and $c$, we always have $\omega_b^+ < \omega_c^-$.  Disobeying this inequality would mean that a scattering process for electrons ($-$ excitations which move forward in time) has the same or smaller energy as some other scattering process for holes ($+$ excitations which move backwards in time) which would imply an electron can scatter into a state below the Fermi energy and/or a hole can scatter into a state above the Fermi energy.  In fact, one expects the opposite: an electron scatters into another electron state above the Fermi energy and emits an excitation (positive energy); a hole scatters into another hole state below the Fermi energy minus some excitation (negative energy).  For an explicit example, within the $GW$ approximation \cite{GWreview1,GWreview2} the inequality is never violated because $\omega_b^+=\epsilon_o-\Omega$ where $\epsilon_o<\mu$ is an occupied state and $\Omega>0$ is some charge excitation such as a plasmon or electron-hole pair while $\omega_u^-=\epsilon_u+\Omega'$ where $\epsilon_u>\mu$ is an unoccupied state and $\Omega'>0$ is some other charge excitation.

The integral of Eq.~(\ref{eq:appendixsnmdef}) has the square of the matrix element, and expanding out the square we have
\begin{multline}
|\bra{n}\Sigma_{xc}(\omega)-U_{xc}\ket{m}|^2  = |r|^2  \\
+ r^*\left[\sum_b \frac{s_b^+}{\omega-\omega_b^+-i\gamma} + \sum_c \frac{s^-_c}{\omega - \omega_c^-+i\gamma}\right]\\
+ r\left[\sum_{b'} \frac{{s_{b'}^+}^*}{\omega-\omega_{b'}^++i\gamma} + \sum_{c'} \frac{{s^-_{c'}}^*}{\omega - \omega_{c'}^--i\gamma}\right]\\
+ \sum_b\sum_{b'} \frac{s_{b}^+{s_{b'}^+}^*}{(\omega-\omega_b^+-i\gamma)(\omega-\omega_{b'}^++i\gamma)}\\
+ \sum_b\sum_{c'} \frac{{s_b^+}{s_{c'}^-}^*}{(\omega-\omega_b^+-i\gamma)(\omega - \omega_{c'}^--i\gamma)}\\
+ \sum_b\sum_c \frac{s_{c}^-{s_{b'}^+}^*}{(\omega-\omega_c^-+i\gamma)(\omega - \omega_{b'}^++i\gamma)}\\
+ \sum_c\sum_{c'} \frac{{s_c^-}{s_{c'}^-}^*}{(\omega-\omega_c^-+i\gamma)(\omega - \omega_{c'}^--i\gamma)}
\label{eq:sigxcuxcnm2}
\end{multline}
We note that only a subset of these terms have poles above the real axis.  Our task it to find the scaling versus $\gamma$ of the contributions from such poles to the integral of Eq.~(\ref{eq:appendixsnmdef}).  To avoid excessively long expressions, we define
\[
g(\omega) \equiv \frac{1}{[(\omega-\epsilon_n)^2+\gamma^2][(\omega-\epsilon_m)^2+\gamma^2]}\,.
\]
The contributions to the integral from the poles of $|\bra{n}\Sigma_{xc}(\omega)-U_{xc}\ket{m}|^2$ are $2\pi i$ times 
\begin{multline*}
r^*\sum_b g(\omega_b^+)s^+_b +  r\sum_{c'} g(\omega_{c'}^-) {s^-_{c'}}^* + \sum_{b,b'}\frac{g(\omega_b^+)s_b^+{s_{b'}^+}^*}{\omega_b^+ - \omega_{b'}^++2i\gamma}\\
+\sum_{b,c'}\frac{\left[g(\omega_b^+)-g(\omega_{c'}^-)\right]s_b^+{s_{c'}^-}^*}{\omega_b^+-\omega_{c'}^-}
+\sum_{c,c'}\frac{g(\omega_{c'}^-){s_c^-}{s_{c'}^-}^*}{\omega_{c'}^--\omega_c^-+2i\gamma}\,.
\end{multline*}
Generally, there is no reason to expect that any of the quasiparticle energies $\epsilon_n$ should precisely equal to any of the excitation energies $\omega_b^+$ or $\omega_c^-$ so that $g(\omega_b^+)$ and $g(\omega_{c'}^-)$ are finite and well behaved as $\gamma\rightarrow0$.  This is rigorously true at low energies for quasiparticle energies close to the Fermi level for a system with an energy gap where the excitations energies $\Omega$ are   greater than or equal to the energy gap.  Therefore, the first two terms  above are not expected to scale strongly with $\gamma$.  The third  ($b,b'$) and fifth  ($c,c'$)  terms scale as $\gamma^{-1}$ for the case when the double sums are discrete and when the real part of their denominators vanish, while the fourth term $(b,c'$) is  finite both by our assumption that $\omega_b^+<\omega_{c}^-$ as well as the fact that the summand is mathematically well behaved as $\omega_b^+\rightarrow\omega_{c'}^-$.  Hence, for discrete sums, the entire contribution has leading scaling behavior $\gamma^{-1}$.  For systems supporting continuous energy bands where the $b$ and $c'$ sums  are continuous integrals, the scaling of the third and fifth terms is reduced to $\gamma^0$ from those bands due to the fact that under an integral
\[
\frac{1}{x+2i\gamma} = P\frac{1}{x} - i \pi \delta(x)
\]
when  $\gamma$ is very small.

In brief, in this appendix we have shown that the contributions to Eq.~(\ref{eq:appendixsnmdef}) that stem from the poles of the numerator $|\bra{n}\Sigma_{xc}(\omega)-U_{xc}\ket{m}|^2$ generically scale as $\gamma^{-1}$ for any discrete poles and as $\gamma^0$ if there are only continuous bands of poles in the self energy (i.e, branch cuts).

\section{Energy change $\Delta F$}
\label{appb}

Here we provide more details for the explicit expression for $\Delta F$ in Eq.~(\ref{eq:deltaFnm}).  For simplicity, let us focus on a single diagonal contribution where $n=m$:
\[
\Delta F_{nn} \equiv -\frac{1}{2}
\int_{-\infty}^\infty  \frac{d\omega}{2\pi i}e^{i\omega 0^+} 
\frac{\bra{n}\Sigma_{xc}(\omega)-U_{xc}\ket{n}^2}{(\omega-\epsilon_n-i\gamma s_n)^2}\,.
\]
The exponential factor in the integrand allows us to turn this into a contour integral by closing the integral contour over the upper complex $\omega$ half plane.  We then obtain two sets of contributions:   residues that  originate from the pole at $\epsilon_n+i\gamma s_n$ which only contribute when $s_n>0$ ({\it i.e.}, $n$ is an occupied state $\epsilon_n<\mu$), and residues originating from the poles of $\Sigma_{xc}(\omega)$ in the numerator.
Inserting the form for $\Sigma_{xc}$ from Eq.~(\ref{eq:sigpoleform}) into the above integral and performing the contour integral, one arrives at a first expression
\begin{multline*}
\Delta F_{nn} = -\theta(\mu-\epsilon_n)\bra{n}\Sigma_{xc}(\epsilon_n)-U_{xc}\ket{n}\bra{n}\frac{d\Sigma_{xc}(\epsilon_n)}{d\omega}\ket{n}\\
- r\sum_b \frac{s^+_b}{(\omega_b^+-\epsilon_n)^2}
+  \sum_b \frac{(s^+_b)^2}{(\omega_b^+-\epsilon_n)^3}\\
- \sum_{b<b'} \frac{s^+_b s^+_{b'}}{\omega_b^+-\omega_{b'}^+}\left[\frac{1}{(\omega^+_b-\epsilon_n)^2} - \frac{1}{(\omega^+_{b'}-\epsilon_n)^2} \right]\\
- \sum_{b,c}\frac{s^+_b s^-_c}{(\omega_b^+-\epsilon_n)^2(\omega_b^+-\omega_c^-)}\,.
\end{multline*}
The first contribution for occupied states comes from the pole of the denominator at $\epsilon_n$ while the remaining terms come from the poles of $\Sigma_{xc}$ at $\omega^+_b$ and $\omega_c^-$.  For unoccupied states $\epsilon_n>\mu$, only the second set of terms contribute.  For occupied states, we can use Eq.~(\ref{eq:deltaFnm}) for the first term and perform some algebra to simplify.  The final results are for an unoccupied state $n=u$ we have
\begin{multline*}
\Delta F_{uu} = 
- r\sum_b \frac{s^+_b}{(\omega_b^+-\epsilon_u)^2}
+  \sum_b \frac{(s^+_b)^2}{(\omega_b^+-\epsilon_u)^3}\\
- \sum_{b<b'} \frac{s^+_b s^+_{b'}}{\omega_b^+-\omega_{b'}^+}\left[\frac{1}{(\omega^+_b-\epsilon_u)^2} - \frac{1}{(\omega^+_{b'}-\epsilon_u)^2} \right]\\
- \sum_{b,c}\frac{s^+_b s^-_c}{(\omega_b^+-\epsilon_u)^2(\omega_b^+-\omega_c^-)}\,.
\end{multline*}
while for an occupied state $n=o$ we have
\begin{multline*}
\Delta F_{oo} = 
r\sum_c \frac{s^-_c}{(\omega_c^- - \epsilon_o)^2}
-\sum_{c,c'} \frac{s^-_c s^-_{c'}}{(\omega_c^--\epsilon_o)(\omega_{c'}^--\epsilon_o)^2}\\
-\sum_{b,c}s^+_b s^-_c\left[\frac{1}{(\omega_b^+-\epsilon_o)(\omega_c^--\epsilon_o)^2} + \frac{1}{(\omega_b^+-\epsilon_o)^2(\omega_c^--\epsilon_o)}\right.\\
\left. + \frac{1}{(\omega_b^+-\epsilon_o)^2(\omega_b^+-\omega_c^-)}\right]
\end{multline*}
Longer expressions with similar structures can be derived for non-diagonal elements $n\ne m$.  However, the main problem is that we have no hint as to how to proceed toward minimizing the magnitude of $\Delta F$ based on such expressions.

\section{Unsuitability of minimizing $\delta F/\delta G$}
\label{appc}

We describe the unfavorable consequences of choosing to minimize $|\delta F/\delta G|$ instead of $|\delta F/\delta \Sigma_t|$.  The objective is to show that the variable $G$ is a poor choice to generate the derivative whose length we aim to minimize.

Beginning with the expression of Eq.~(\ref{eq:dFdG}) for $\delta F/\delta G(\omega)$ and evaluating it for non-interacting Green's functions $G=G_0$, we follow the steps in Section~\ref{sec:gradient} to find the derivative
\[
\mathcal{D}_0(\omega) \equiv 2\pi i \frac{\delta F}{\delta G(\omega)}\Big|_{G=G_0} = \Sigma_{xc}(\omega)-U_{xc}\,,
\]
the squared length to be minimized 
\begin{equation*}
||\mathcal{D}_0||^2 =  \int_{-\infty}^\infty d\omega \, tr\{ (\Sigma_{xc}(\omega)-U_{xc})^\dag (\Sigma_{xc}(\omega)-U_{xc})\}\,,
\end{equation*}
and an explicit expression in the non-interacting eigenbasis
\begin{equation*}
||\mathcal{D}_0||^2 = \int_{-\infty}^\infty d\omega \, \sum_{n,m} |\bra{n}\Sigma_{xc}(\omega)-U_{xc}\ket{m}|^2\,.
\end{equation*}
Using the expression for the squared matrix element in Eq.~(\ref{eq:sigxcuxcnm2}) and performing the $\omega$ integral, we find
\begin{multline*}
||\mathcal{D}_0||^2 = \sum_{n,m}\left\{
|r|^2\left(\int_{-\infty}^\infty d\omega\right)\right.\\
 + 2\pi Im \left( r\sum_b {s_{b}^+}^* + r^*\sum_c {s_{c}^-}\right)\\
\left. + \sum_{b,b'} \frac{s_b^+{s_{b'}^+}^*}{\omega_b^+ - \omega_{b'}^++2i\gamma} + \sum_{c,c'}\frac{{s_c^-}{s_{c'}^-}^*}{\omega_{c'}^--\omega_c^-+2i\gamma}\right\}\,,
\end{multline*}
remembering that $r=\bra{n} \Sigma_x - U_{xc}\ket{m}$ is the static part of the matrix element involving the static and nonlocal bare Fock operator $\Sigma_x$ and $U_{xc}$.

Unlike the situation in the main text where the squared length of the derivative is always finite for any $\gamma>0$ ({\it i.e.}, it is a regularized expression since $\gamma$ acts as an infrared cutoff), the above expression for $||\mathcal{D}_0||^2$ is manifestly infinite and nonsensical unless $r=0$ for all choices of $n=m$.  The choice $r=0$ for all $n,m$ means $\Sigma_x=U_{xc}$ or that our non-interacting $G_0$ must correspond to the Hartree-Fock one {\it regardless} of the level of exchange and correlation we decide to include in the exchange-correlation functional $\Phi_{xc}$ or the self-energy $\Sigma_{xc}$.  

This is a very poor situation indeed: by choosing to minimize $|\delta F/\delta G|$, we are forced to adopt the Hartree-Fock solution for the non-interacting Green's function $G_0$ not because it necessarily represents an optimum choice but simply to avoid literal infinities in our mathematical description.  By contrast, minimizing $|\delta F/\delta \Sigma_t|$ yields finite answers when $\gamma>0$ which permits us to proceed in our analysis and  find an optimum choice of $G_0$.

\section{Behavior of self-energy as $\omega\rightarrow\infty$}
\label{appd}

If we assume that equation Eq.~(\ref{eq:sigxcpoles}) is true and that the pole energies $\xi_a$ are finite, then it is clear that $\lim_{\omega\rightarrow\infty}\Sigma_{xc}(\omega)$ is finite and equal to $\Sigma_x$.  Separately, for such forms of self-energy, any   excitation energies such as $\xi_a$ will be finite for a finite physical system solved in a finite basis set.

\red{For the more general case, we can show that $\Sigma_{xc}(\omega)$ grows at most sublinearly with $\omega$ as $\omega\rightarrow\infty$ without any assumptions about the functional form of $\Sigma_{xc}(\omega)$.}  This fact follows directly from the Lehmann representation of the  zero-temperature, time-ordered, many-body Green's function for an $N$-electron system:
\[
G(x,x',\omega) = \sum_s \frac{f_s(x)f_s(x')^*}{\omega-e_s+i0^+sgn(e_s-\mu)}
\]
where
\[
f_s(x) = \langle N,0|\hat \psi(x)| N+1,s \rangle \ \mbox{if } e_s>\mu \ ; \ 
f_s(x) = \langle N-1,s|\hat \psi(x)| N,0 \rangle \ \mbox{if } e_s<\mu
\]
and $\ket{M,s}$ labels the exact many-body eigenstate with $M$ electrons and energy $E_{M,s}$ where $s=0$ is the ground state.  For electron-like excitations $e_s=E_{s,N+1}-E_{N,0}>\mu$ while for hole-like excitations $e_s=E_{N,0}-E_{N-1,s}<\mu$.  The electron annihilating field operator is $\hat\psi(x)$.  When we send $\omega\rightarrow\infty$, completeness of the eigenbasis and the canonical commutation relation $\{\hat\psi(x),\hat\psi^\dag(x')\}=\delta(x,x')$ lead to
\[
\lim_{\omega\rightarrow\infty} G(x,x',\omega) = \frac{\delta(x-x')}{\omega}\,.
\]
\red{
In this limit, one can easily invert the $G(\omega)$ matrix to find
 \[
\lim_{\omega\rightarrow\infty} G(\omega)^{-1} = \omega I\,.
\]
}
\red{Comparing this with the Dyson Eq.~(\ref{eq:traddyson}) shows that $\Sigma_{xc}(\omega)/\omega$ must vanish as $\omega\rightarrow\infty$.  Hence, at most, $\Sigma_{xc}(\omega)$ grows sublinearly with $\omega$ as $\omega\rightarrow\infty$.}

\section{Choosing largest $Z$ in the multi-pole and multi-band cases}
\label{appe}

Here we describe analytical results and bounds for optimizing the length of the gradient in situations where there are multiple solutions to the QS equations.  Once the self-consistency conditions in Eq.~(\ref{eq:qsgw}) are obeyed, since we work in the diagonal basis of the non-interaction Hamiltonian $H_0$ determined by this condition, the information content of the non-interacting eigenvalue equation is given by the diagonal QS energy-only Dyson-type equation 
\[
(H_0)_{n,n} = \epsilon_n = (T+U_{ion}+\phi_H+\Sigma_{xc}(\epsilon_n) )_{n,n}\,.
\]
For a self-energy given by the form of Eq.~(\ref{eq:sigxcpoles}), which we write more explicitly here for the case of $p$ poles with real excitation energies $\xi_a$,
\[
\Sigma_{xc}(\omega) = \Sigma_x + \sum_{a=1}^p \frac{\sigma^a}{\omega-\xi_a \pm i\gamma} \,.
\]
the above energy-only QS condition requires finding the roots $\epsilon_n$ of a polynomial of order $p+1$ (we are only interested in the real-valued solutions to describe a real-valued non-interacting energy $\epsilon_n$).  The question is which of the possible $p+1$ solutions has the shortest gradient of the LW functional.  We analyze the single-band and multi-band cases separately.  We will be assuming time-reversal symmetry and making $\gamma$ very small to simplify the mathematics.

\subsection{Single-band case}

Here we only have a single band $n$ to worry about, and the question is which solution $\epsilon_n$ minimizes 
\[
S_{nn} = \int_{-\infty}^\infty d\omega \frac{|\Delta(\omega)_{nn}|^2}{[(\omega-\epsilon_n)^2+\gamma^2]^2}
\]
where we define the shorthand
\[
\Delta(\omega) = \Sigma_{xc}(\omega)-U_{xc}
\]
and the QS condition means $\Delta(\epsilon_n)_{nn} = 0$.  

One contribution to $S_{nn}$ for very small $\gamma$ comes from the pole at $\omega=\epsilon_n$: near this pole, $\Delta(\omega)_{nn}\approx \Sigma_{xc}'(\epsilon_n)_{nn}\cdot(\omega-\epsilon_n\pm i\gamma)$, and plugging this in permits analytical evaluation of the integrals for $\gamma$ very small (using $\int dx\, 1/(1+x^2)^2=\int dx\, x^2/(1+x^2)^2=\pi/2$).  The second set of contributions to $S_{nn}$ for very small $\gamma$ come from the poles of the self-energy near $\xi_a$: for those parts of the integral, the denominator of the integrand is essentially constant and the numerator has a classic Lorentzian form which is easily integrated.  We  arrive at
\[
S_{nn} = \frac{\pi}{\gamma}\cdot\left( |\Sigma'_{xc}(\epsilon_n)_{nn}|^2 + \sum_{a=1}^p \frac{|\sigma^a_{nn}|^2}{(\epsilon_n-\xi_a)^4}\right)\,.
\]
Plugging an explicit form for $\Sigma_{xc}'(\epsilon_n)$ gives
\[
S_{nn} = \frac{\pi}{\gamma}\cdot\left( \left|\sum_{a=1}^p \frac{\sigma^a_{nn}}{(\epsilon_n-\xi_a)^2}\right|^2 + \sum_{a=1}^p \frac{|\sigma^a_{nn}|^2}{(\epsilon_n-\xi_a)^4}\right)\,.
\]

If we have a single pole ($p=1$), then both terms are equal and $S_{nn} = (2\pi/\gamma)|\Sigma'_{xc}(\epsilon_n)_{nn}|^2$ so that the solution with smallest $|\Sigma'_{xc}(\epsilon_n)_{nn}|$ (i.e., largest $Z$) will have the smallest gradient.

For multiple poles ($p>1$), we will be assuming that the matrices $\sigma^a$ are positive definite and Hermitian (they are such in the $GW$ approximation regardless of whether we use the RPA approximation for $W$ or the exact $W$~\cite{myPRB}).  In addition, this is a sensible assumption since it guarantees the diagonal elements of the self energy to have decreasing slopes with $\omega$ so quasiparticle weights $Z=1/(1-d\Sigma_{xc}/d \omega)$ fall in the physical range $0\le Z\le1$.  Thus, $\sigma^a_{nn}=\bra{n}\sigma^a\ket{n}\ge 0$ will be assumed.

Given that $\sigma^a_{nn}\ge0$, the above expression for $S_{nn}$ has sums of $p$ positive quantities being squared separately or squared after summation, and we can bound the values of the sum.  We define a vector $\vec v_n$ with $p$ real-valued component $v_{an} = \sigma^a_{nn}/(\epsilon_n-\xi_a)^2$.  Then we can write $S_{nn}$ as 
\[
\gamma S_{nn}/\pi = \left( \vec v_n \cdot \vec d\,\right)^2 + \Vert \vec v_n \Vert^2 \equiv P_n + Q_n
\]
where $\vec d = (1,1,\ldots,1)^T$.  The first term $P_n$ is the square of the projection $\vec v_n$ onto $\vec d$, and the second term $Q_n$ is the Euclidean squared length of $\vec v_n$.  If we fix $P_n$, we are forcing $\vec v_n$ to lie on a hyperplane perpendicular to $\vec d$.  For fixed $P_n$, the minimum value of $Q_n$ occurs when $\vec v_n$ is parallel to $\vec d$ (i.e, all components of $\vec v_n$ are equal), in which case $Q_n/P_n=1/p$.  For fixed $P_n$, the maximum value of $Q_n$ happens on the extreme corners of the hyperplane in the allowed region $v_{an}\ge0$, which means there is only one non-zero component of $\vec v_n$; in this case, $Q_n/P_n=1$.  Since $P_n=|\Sigma'_{xc}(\epsilon_n)_{nn}|^2$, we have the bound
\[
(1+1/p)\cdot|\Sigma'_{xc}(\epsilon_n)_{nn}|^2 \, \le \, \gamma S_{nn}/\pi\,  \le \, 2|\Sigma'_{xc}(\epsilon_n)_{nn}|^2\,.
\]
$S_{nn}$ is clearly controlled directly by $|\Sigma'_{xc}(\epsilon_n)_{nn}|^2$, so all else being equal, smaller $|\Sigma'_{xc}(\epsilon_n)_{nn}|$  gives smaller $S_{nn}$ and hence shorter gradients.  Specifically, if any two solutions have a ratio of their $\Sigma'_{xc}(\epsilon_n)_{nn}$ larger than $\sqrt{2}$, then choosing the smaller $\Sigma'_{xc}(\epsilon_n)_{nn}$ will definitely lead to a shorter gradient.  It is difficult to say much more  {\it a priori} without knowing more about the distribution of the $p$ components $v_{an}$.

\subsection{Multi-band case}

For multi-band cases, one must also consider the minimization of off-diagonal contributions  $S_{nm}$ to the length of the gradient for $n\ne m$.  We will be focusing on the non-degenerate case $\epsilon_n\ne\epsilon_m$ since, based on the discussion in the matin text, the degenerate case reverts to the diagonal $n=m$ case.  Reproducing Eq.~(\ref{eq:Snmoffdiag}), we have
\[
S_{nm} = \int_{-\infty}^\infty d\omega \frac{|(\Sigma_{xc}(\omega)-U_{xc})_{nm}|^2}
{[(\omega-\epsilon_n)^2+\gamma^2][(\omega-\epsilon_m)^2+\gamma^2]}\,,
\]
where $\Delta(\omega) = \Sigma_{xc}(\omega)-U_{xc}$.  
For small $\gamma$, we have contributions from the two poles in the denominators (see Eq.~(\ref{eq:nmpolescontrib})) and contributions from the poles of $\Sigma_{xc}(\omega)$. Together they give
\[
\gamma S_{nm}/\pi =  \frac{| (\Sigma_{xc}(\epsilon_n)-U_{xc})_{nm}|^2 + |(\Sigma_{xc}(\epsilon_m)-U_{xc})_{nm}|^2}{(\epsilon_n-\epsilon_m)^2} + \sum_{a=1}^p \frac{|\sigma^a_{nm}|^2}{(\epsilon_n-\xi_a)^2(\epsilon_m-\xi_a)^2}\,.
\] 
The QS condition of Eq.~(\ref{eq:qsgw}) optimizes the first term in the sum above.  Plugging in the optimal choice, and then performing some algebraic rearrangements using the fact that $(U_{xc})_{nm}$ is real by time-reversal invariance, we have
\begin{multline*}
\gamma S_{nm}/\pi = \frac{(Im\Sigma_{xc}(\epsilon_n)_{nm})^2+(Im\Sigma_{xc}(\epsilon_m)_{nm})^2+ \left[Re\Sigma_{xc}(\epsilon_n)_{nm} - Re\Sigma_{xc}(\epsilon_m)_{nm}\right]^2/2}{(\epsilon_n-\epsilon_m)^2 }\\
+ \sum_{a=1}^p \frac{|\sigma^a_{nm}|^2}{(\epsilon_n-\xi_a)^2(\epsilon_m-\xi_a)^2}\,.
\end{multline*}

As discussed in the main text, the optimization over $U_{xc}$ can not help us with the imaginary parts since $U_{xc}$ is Hermitian: we are ``stuck'' with whatever these values may be.  In what follows, we will ignore the contributions from the imaginary part by assuming they are either zero for the states of interest or small.   Dropping these terms and plugging in the explicit formula for the self-energy, we arrive at
\[
\gamma S_{nm}/\pi = \frac{1}{2}\left(\sum_{a=1}^p\frac{Re\,\sigma^a_{nm}}{(\epsilon_n-\xi_a)(\epsilon_m-\xi_a)}\right)^2
+ \sum_{a=1}^p \frac{|\sigma^a_{nm}|^2}{(\epsilon_n-\xi_a)^2(\epsilon_m-\xi_a)^2}\,.
\]
Our assumptions of time-reversal invariance and zero part of the imaginary part of the self-energy mean that $\Sigma(r,r',\omega)$ will be real valued: hence the matrix elements $\sigma^a_{nm}$ are also real valued.  Thus we have
\[
\gamma S_{nm}/\pi = \frac{1}{2}\left(\sum_{a=1}^p\frac{\sigma^a_{nm}}{(\epsilon_n-\xi_a)(\epsilon_m-\xi_a)}\right)^2
+ \sum_{a=1}^p \frac{(\sigma^a_{nm})^2}{(\epsilon_n-\xi_a)^2(\epsilon_m-\xi_a)^2}\equiv \frac{1}{2}P_{nm} + Q_{nm}\,.
\]
To put bounds on this expression, we will require all contributions to be positive inside the squared term for $P_{nm}$.  Taking absolute values, we have
\[
P_{nm} \le \left(\sum_{a=1}^p\frac{|\sigma^a_{nm}|}{|(\epsilon_n-\xi_a)(\epsilon_m-\xi_a)|}\right)^2\,.
\]

At this point, we invoke the Cauchy-Schwarz inequality which says that $(\sigma^a_{nm})^2\le \sigma^a_{nn}\sigma^a_{mm}$: this following from defining an inner product with the matrix $\sigma^a$ as the metric, and the positive definiteness of $\sigma^a$ leads to this bound.  So 
\[
P_{nm} \le \left(\sum_{a=1}^p\frac{\sqrt{\sigma^a_{nn}\sigma^a_{mm}}}{|\epsilon_n-\xi_a|\cdot|\epsilon_m-\xi_a|}\right)^2 = \left(\sum_{a=1}^p \sqrt{v_{an}}\sqrt{v_{am}}\right)^2
\]
where we have reintroduced the vectors $\vec v_n$ from the previous sections.  Viewing the last term as an inner product and applying Cauchy-Schwarz this time to vectors with components $\sqrt{v_{an}}$, we have
\[
P_{nm} \le \sqrt{P_nP_m}
\]
where $P_{n}$ and $Q_n$ were defined in the previous section.  Similar logic shows that
\[
Q_{nm} \le \sqrt{Q_nQ_m}\,.
\]
Since we know how $P_n$ and $Q_n$ are related, we can put an upper bound on $S_{nm}$ which is
\[
S_{nm} \le \frac{3}{2}\sqrt{P_nP_m} = | \Sigma'_{xc}(\epsilon_n)_{nn} |\cdot| \Sigma'_{xc}(\epsilon_m)_{mm} |\,.
\]
Again, choosing the solution for each state to have the smallest $|\Sigma'_{xc}(\epsilon_n)_{nn}|=\sqrt{P_n}$ will ensure $S_{nm}$ is as small as possible.

\section*{References}
\bibliographystyle{unsrt}
\bibliography{qpscgradient-lifetime}

\begin{thebibliography}{10}
\bibitem{higherordernote}
In practice, we will not know the extremizing $\Sigma_{xc}$ in
  Eq.~(\ref{eq:deltaF}) since we are not actually solving the Dyson equation
  for the extremizing $\bar G$. Instead, if one is using non-interacting
  Green's function $G_0$, one has access to an approximate self-energy that is
  evaluated at that $G_0$, namely $\Sigma_{xc}^0=\delta\Phi_{xc}[G]/\delta
  G|_{G=G_0}$. However, since we are series expanding $\Delta F$ about the
  extremum, the error in $\Sigma_{xc}^0$ is linear in $\Sigma_{xc}-U_{xc}$ and
  thus contributes to cubic and higher order terms in $\Delta F$ so the
  quadratic expression of Eq.~(\ref{eq:deltaF}) remains correct at quadratic
  order.
\end{thebibliography}

\end{document}